\documentclass[reprint,rmp]{revtex4-2}
\usepackage{verbatim}
\usepackage{physics,amsmath,amssymb}
\usepackage{booktabs}
\usepackage{color}
\usepackage{soul}

\usepackage{placeins} 
\usepackage{url}
\usepackage[hyperindex,breaklinks]{hyperref}
\hypersetup{linkcolor=blue, citecolor=blue, colorlinks=true}

\usepackage{capt-of}
\usepackage{graphicx}
\usepackage{dcolumn}
\usepackage{bm}

\usepackage[margin=1.5cm]{geometry}

\begin{document} 

\newtheorem{theorem}{Theorem}
\newtheorem{corollary}{Corollary}
\newtheorem{definition}{Definition}
\newtheorem{example}{Example}
\newtheorem{lemma}{Lemma}
\newtheorem{proposition}{Proposition}
\newtheorem{remark}{Remark}

\title{
Chaos and Parrondo's paradox: An overview
}

\author{Marcelo A. Pires$^{1}$}
\thanks{piresma@cbpf.br}

\author{Erveton P. Pinto$^{2}$}
\thanks{erveton@unifap.br}

\author{Jose S. Cánovas$^{3}$}
\thanks{jose.canovas@upct.es}

\author{Silvio M. Duarte Queir\'os$^{1,4}$}
\thanks{sdqueiro@cbpf.br}

\affiliation{$^{1}$Centro Brasileiro de Pesquisas F\'isicas, Rio de Janeiro/RJ, 22290-180, Brazil 
\\ %
$^{2}$Departamento de Ciências Exatas e Tecnológicas, Universidade Federal do Amapá, Macapá/AP, 68903-419, Brazil
\\ 
$^{3}$Departamento de Matemática Aplicada y Estadística, Universidad Politécnica de Cartagena, C/ Doctor Fleming sn, 30202, Cartagena, Spain
\\
$^{4}$National Institute of Science and Technology for Complex Systems, Rio de Janeiro/RJ, 22290-180, Brazil
}

\begin{abstract}
Parrondo's paradox  (PP) is a fundamental principle in nonlinear science where the alternation of individually losing strategies leads to a  winning outcome. 
In this topical review, we provide the first  systematic panorama of the synergy between PP and chaos. We observe a bidirectional connection between the two areas. 
The first direction is the translation of PP into  the interplay between Order and Chaos through either Chaos + Chaos $\to$ Order (CCO) or Order + Order $\to$ Chaos (OOC). In this vein, many quantifiers, such as Lyapunov Exponents, $\lambda$, and entropic measures, are used.
Second, we note that chaos can be used to engineer switching protocols that can lead to nontrivial effects in diverse PP cases. 
Our review clarifies the universality of PP and highlights its robust theoretical and practical applications across several areas of science and technology. 
Finally, we delineate key open questions, emphasizing the unresolved theoretical limits, the role of high-dimensional maps and continuous flows, and the critical need for more experimental verification of the dynamic PP in chaotic systems.
For completeness, we also provide a full Python code that allows the reader to observe the many facets of the PP.
\end{abstract}


\maketitle

\tableofcontents

\section{Introduction}

\begin{table*}[!htb]
\caption{Types of  Parrondo's paradox (PP)   related to chaos theory when iterating two suitable maps $f$ and $g$. Note that $\lambda$ is the Lyapunov exponent and $h$ is the topological entropy.}
\label{tab:v2}
\begin{tabular}{@{}l w{c}{6cm} w{c}{6cm}@{}}
\toprule
\textbf{Type} & \textbf{Chaos + Chaos $\to$ Order (CCO)} & \textbf{Order + Order $\to$ Chaos (OOC)} \\
\midrule
\textit{Physical  } & 
$\lambda(f)>0, \ \lambda(g) > 0 \implies \lambda(f,g) < 0$ & 
$\lambda(f)<0, \ \lambda(g) < 0 \implies \lambda(f,g) > 0$ \\
\midrule
\textit{Topological} & 
$h(f)>0, \ h(g) > 0 \implies h(f,g) = 0$ & 
$h(f) = h(g) = 0 \implies h(f,g) > 0$ \\
\midrule
\textbf{Outcome} & 
Suppression of chaos via switching & 
Emergence of chaos via switching \\
\bottomrule
\end{tabular}
\label{tab:parrondo_types}
\end{table*}

\begin{figure}[!htb]
    \centering
    \includegraphics[width=0.99\linewidth]{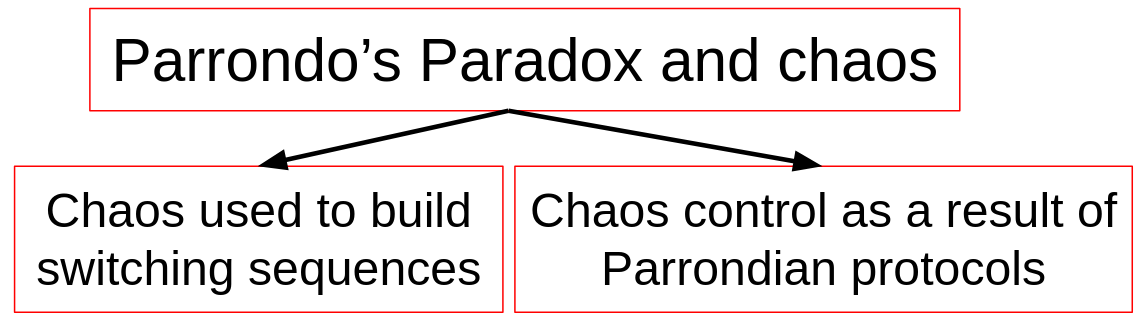}
    \caption{Flowchart with the bidirectional connections between the  PP  and chaos theory. } 
    \label{fig:flowchart}
\end{figure}

Parrondo's paradox (PP) is a highly compelling and counterintuitive principle in nonlinear science, based on the paradoxical defiance of linear summation. It demonstrates that the sequential alternation of two or more distinct processes -- each individually tending toward a `losing' or entropically-disadvantageous outcome -- can unexpectedly culminate in a collective `winning' net gain. This mechano-conceptual bedrock finds its lineage in the physics of non-equilibrium systems, specifically the Flashing Brownian Ratchet model. This physical analogue establishes the non-commutative prerequisite: the crucial requirement for \textit{state dependence} or the \textit{asymmetry} induced by switching, thereby enabling the combined dynamics to nonlinearly evade the entropic fate of its components.

The transition of the PP from the domain of stochastic processes to the precise, deterministic architecture of dynamical systems marks a crucial theoretical step. Here, the game-theoretic concepts are transmuted and rigorously mapped onto the phase variable of the system, i.e., losing is the attainment of \textit{Order} or \textit{dynamically regular} behavior -- e.g., convergence to a periodic orbit characterized by a negative Lyapunov exponent, $\lambda<0$ --, while winning is the unleashing of \textit{Chaos} or \textit{dynamically complex} behavior -- e.g., exponential instability, $\lambda>0$). This translation yields the dual, fundamental facets of the phenomenon, as shown in table~\ref{tab:parrondo_types}; on the one hand, the suppression of exponential instability, the \textbf{Chaos + Chaos $\to$ Order} (CCO) paradox, and on the other hand the counter-intuitive emergence of complexity from simplicity, the \textbf{Order + Order $\to$ Chaos} (OOC) paradox.

Notwithstanding this growing body of knowledge, a systematic synthesis that rigorously bridges the PP phenomenon with the full conceptual expanse of chaos theory remains a lacuna in the literature. 
We address this lacuna from a bidirectional point of view, as illustrated in Fig.~\ref{fig:flowchart}. 
Our objective unfolds threefold: first, to craft a rigorous comparative framework that elucidates the paradox's dichotomy of perception—the difference between \textit{observable chaos} (Lyapunov exponents) and the \textit{topological chaos} (topological entropy); second, to dissect the fundamental bidirectional connection between PP and chaos, covering the utilization of complex dynamics for superior switching protocols and the application of Parrondian logic for the control of instability in high-dimensional systems; and finally, to synthesize the diverse and robust interdisciplinary applications of the dynamic PP, ranging from quantum systems and engineering optimization to machine learning and ecological modeling, thereby revealing its profound utility as a universal lever for manipulating the geometry of complexity.

As a by-product, aiming to make the review worthwhile to both the Physics and Mathematics communities, we endeavor to play down the intrinsic tension between definitions of chaos based on $\lambda > 0$ (Physics) and the entropic/topological definitions (Mathematics) that arises from the distinction between observational utility and structural rigor.
In most cases, we assume that the evaluation of a positive Lyapunov exponent, $\lambda > 0$, in the study of the Chaotic Parrondo Paradox is sufficient to characterize chaos, provided that the argument is formally anchored in two premises: {\it i)} $\lambda > 0$ serves as a direct measure of metric entropy, $h_{\mu}$, through Pesin's identity, $h_{\mu} = \sum \lambda^+$, by which we impose an ergodic connectivity, and {\it ii)} the model behaves as a typical or observable dynamical system which allows us to invoke Sander \& Yorke's Conjecture 2~\cite{sander2015many}, ensuring that analytical chaos -- related to the Lyapunov exponent -- implies structural and topological chaos.
By integrating these justifications, we maintain the clarity and quantification required by Physics while satisfying the standards of Mathematical rigor and the theory of dynamical systems.

\section{General aspects}

Despite the existing scholarly literature on PP~\cite{harmer2002review,abbott2010asymmetry,lai2020parrondo,cheong2019paradoxical,wen2024parrondo}, a comprehensive elucidation of the body of knowledge linking this phenomenon to chaos theory has not been provided. This work aims to address this lacuna.

\begin{figure}
    \centering
    \includegraphics[width=0.89\linewidth]{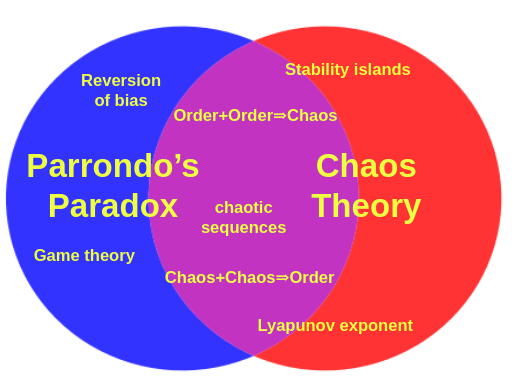}
    \caption{Venn diagram showing some key points in the bridges between the  PP  and chaos theory. } 
    \label{fig:diagram_venn}
\end{figure}

\subsection{Topics that are not the focus of this review}

Our review focuses on the intersection between chaos and the parrondian phenomena as depicted in 
Fig.\ref{fig:diagram_venn}. Thus, although we discuss aspects of chaos control, this topic is not our focus. In the same vein, we do not cover all aspects of the PP that do
not have a direct connection with chaos.

\subsection{An Overview with VOSviewer}

\begin{figure*}
    \centering
    \includegraphics[width=0.99\linewidth]{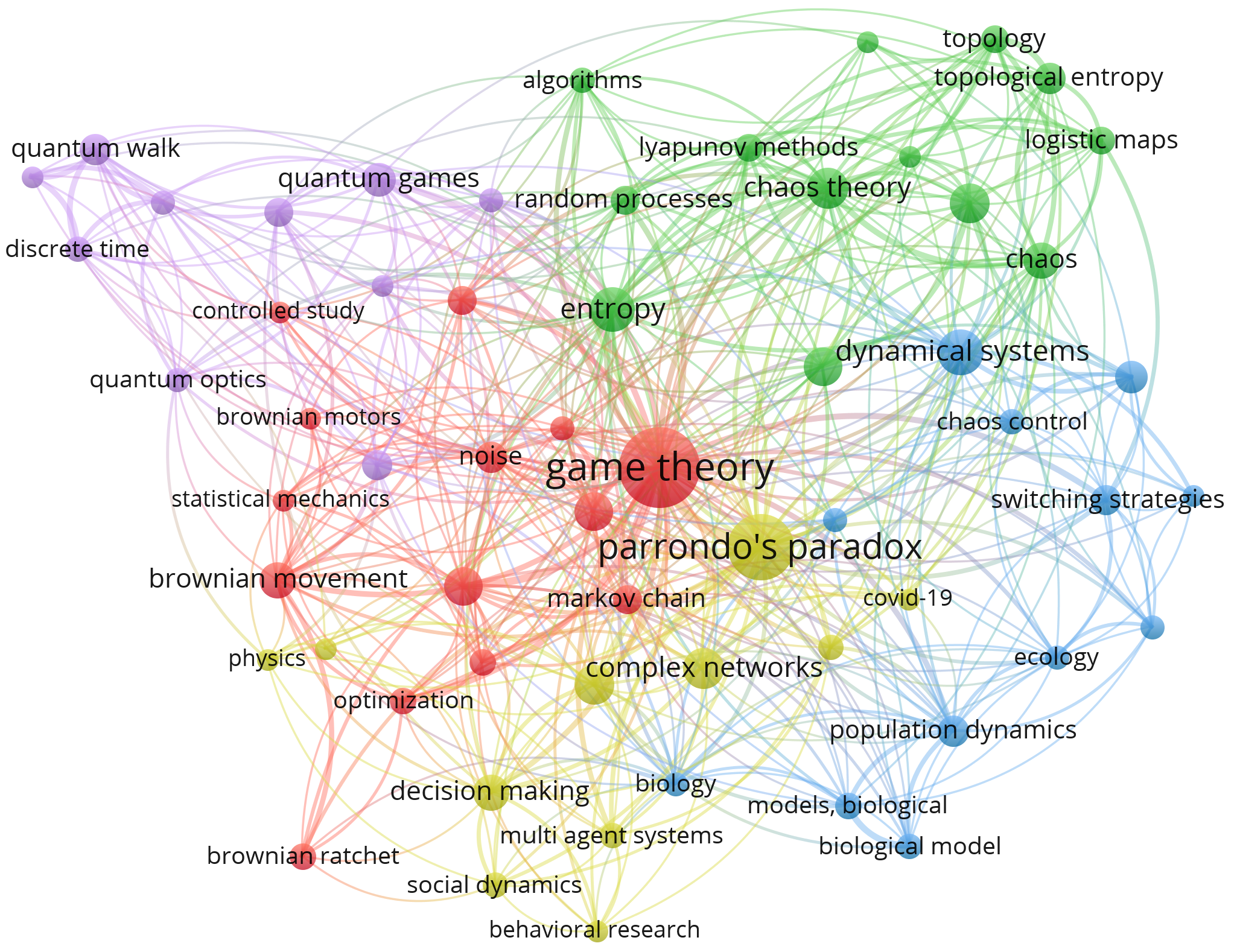}
    \caption{The VOSviewer map presented visually organizes research topics related to  PP  combined with chaos into distinct clusters, each represented by a different color. This clustering highlights the diverse applications and interconnectedness across various scientific disciplines.} 
    \label{fig:VOSviewer}
\end{figure*}

\begin{figure*}
    \centering
    \includegraphics[width=0.99\linewidth]{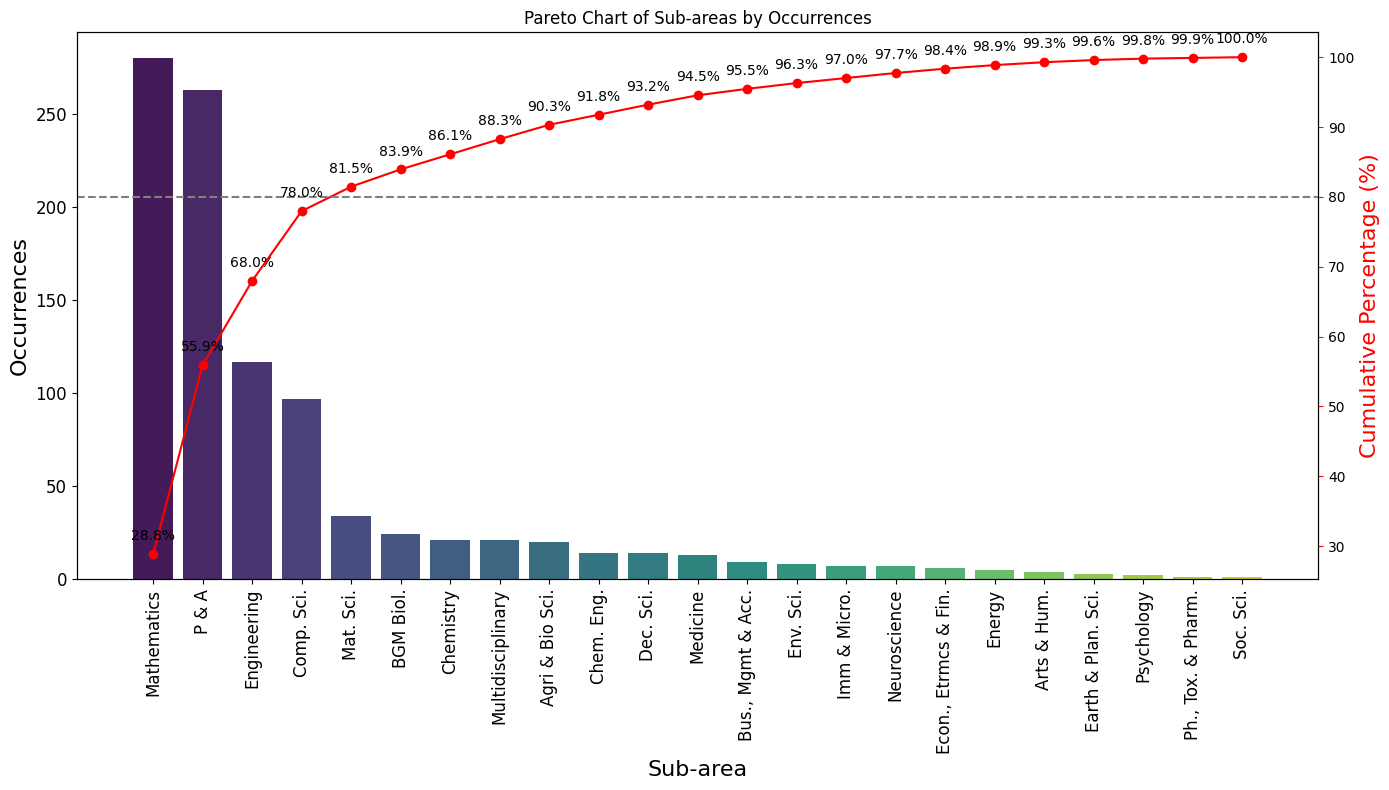}
    \caption{Pareto diagram of the main areas of knowledge that employ  PP  and Chaos in a combined way. P \& A: Physics and Astronomy; Comp. Sci.: Computer Science; Mat. Sci.: Materials Science; BGM Biol.: Biochemistry, Genetics and Molecular Biology; Agri \& Bio Sci.: Agricultural and Biological Sciences; Chem. Eng.: Chemical Engineering; Dec. Sci.: Decision Sciences; Bus., Mgmt \& Acc.: Business, Management and Accounting; Env. Sci.: Environmental Science; Imm \& Micro.: Immunology and Microbiology; Econ., Etrmcs \& Fin.: Economics, Econometrics and Finance; Arts \& Hum.: Arts and Humanities; Earth \& Plan. Sci.: Earth and Planetary Sciences; Ph., Tox. \& Pharm.: Pharmacology, Toxicology and Pharmaceutics; Soc. Sci.: Social Sciences.} 
    \label{fig:Pareto_chart}
\end{figure*}

To analyze the applications that combine  PP  and Chaos, we employed VOSviewer~\cite{van2010software}, a widely used free software tool for constructing and visualizing bibliometric networks. The primary purpose of using VOSviewer was to identify and map the landscape of this research topic. The networks generated by VOSviewer are based on co-occurrence data for keywords extracted from a Scopus dataset. The Scopus database was used because it presented the largest quantity and diversity of works identified through the search terms `PP' and `Chaos'.

To generate the keyword co-occurrence network, we chose the `Full counting' method and set the minimum number of occurrences for a keyword to 7. We also filtered out irrelevant words such as `article', `review', and others. After preprocessing, the network yielded 60 items, 5 clusters, and 518 links. Each cluster was assigned a unique color for visual differentiation, allowing for an intuitive understanding of the thematic subdivisions, as shown in Figure~\ref{fig:VOSviewer}. This way, items that frequently occur together are placed closer together, forming clusters that represent related research topics or application areas.

The red group highlights the topic's relevance in mathematical sciences, with a focus on Game Theory, which is expected given the origin of  PP. The purple group emphasizes the potential application of this tool to problems in Quantum Theory, such as quantum optics and quantum walks. The green group corroborates the relationship between PP and Chaos Theory, with applications ranging from mathematical models, such as the Logistic Map, to practical applications, such as cryptography. The blue and yellow groups highlight the interdisciplinary nature of the topic by focusing on applications to complex dynamic systems across a wide range of fields, including ecology, social and population dynamics, financial markets, biological models, and others.

To complement the VOSviewer analysis, Figure~\ref{fig:Pareto_chart} illustrates the Pareto chart of major knowledge areas employing  PP  and Chaos in a combined way. The results show interdisciplinary applications across various areas of knowledge, with an emphasis on mathematics, the physical sciences, engineering, and computer science, which together account for almost $80\%$ of publications during 2002 to September 2025. Furthermore, our results reveal that there are potential areas that are little explored, such as medicine, environmental and social sciences, and finance, among others.

\subsection{Historical Remarks}\label{sec:refs}

In 1999, the  PP  was formally introduced in the realm of stochastic games~\cite{harmer1999losing}. In subsequent years, this model was  considered within the framework of history-dependent games~\cite{parrondo2000new}.

Around 2002, initial scientific efforts began to explicitly link the PP to chaos~\cite{bucolo2002does,kocarev2002lyapunov}. Also in 2002, the first review of the  PP was presented~\cite{harmer2002review}. This review was later updated in 2010~\cite{abbott2010asymmetry}.

In 2006, in~\cite{canovas2006dynamic}, the PP is studied in general for continuous interval maps. The study includes topological notions as topological entropy, Li-Yorke chaos, and periodic structure of the maps, among others.

In 2013, in~\cite{canovas2013revisiting}, both versions of the PP, ``chaos+chaos$\to$order'' and ``order+order$\to$chaos'' was considered for the logistic family. Chaos was considered in a topological setting, via topological entropy, and in an observable way, via Lyapunov exponents. Still, it is unknown whether the paradox ``chaos+chaos$\to$order'' is possible in the topological sense. As far as we are aware, the paradox `chaos+chaos$\to$order' with Lyapunov exponents is only possible when the topological entropy is positive.
 
The PP has also been analyzed in dynamics beyond 1D models. The paradox was studied in high-dimensional discrete systems, particularly with respect to the stability of their fixed points~\cite{cat1}. Furthermore, continuous-time chaotic models were explored in~\cite{danca2013convergence}. The triple connection among chaos, PP, and quantum systems was first introduced in 2021~\cite{panda2021order,lai2021chaotic}.

Until now, research on the connections between Parrondo’s paradox and chaos has remained fragmented; this review addresses this gap by providing the first broad panorama of this relationship.

\section{Basic background}

\subsection{Parrondian games}

The Parrondian games can be introduced in different ways~\cite{harmer1999losing,harmer2002review,abbott2010asymmetry}, and here we follow the simplified version presented in Section 7.5 of the book~\cite{mitzenmacher2017probability} that has a minimal solvable framework for the  PP  in terms of a  Gambler's Ruin-style Markov chain. In this way, the reader can understand the main ideas without undue concern about complex details.
 
\vskip 0.3cm 
\noindent \textbf{Rules}   \\ \normalsize

Consider a PP through a simplified Gambler's Ruin framework with capital bounds at $C=-3$ and $C=3$. Then, the model operates on a finite state space $\{-3,-2,-1,0,1,2,3\}$ with absorbing boundaries.

The Gambler's Ruin approach simplifies the explanation of PP by focusing on absorption probabilities ($z_j$) rather than long-term averages. This formulation converts the paradox into a simple question: What is the probability of reaching $C=3$ before $C=-3$?

\vskip 0.3cm 
\noindent \textbf{Game A}   \\ \normalsize

Game A is a simple game:
\begin{itemize}
    \item Head (gain $C=1$) with probability $p^{a} = 0.49$
    \item Tail (lose $C=1$) with probability $q^{a} = 0.51$
\end{itemize}

We model Game A using a Markov chain to determine the absorption probability $z_j$ (the probability of reaching $C=3$ before $C=-3$ from state $j$).
The system of equations is:
\begin{align}
z_{-3} &= 0, \quad z_3 = 1, \\
z_j &= q^{a} z_{j-1} + p^{a} z_{j+1} \quad \text{for } j \in \{-2, -1, 0, 1, 2\}
\end{align}
The solution for $z_0$ is:
\begin{align}
z_0^A &= \frac{(\frac{q^{a}}{p^{a}})^3 - 1}{(\frac{q^{a}}{p^{a}})^6 - 1} \quad \text{for } p^a \ne q^a
\end{align}
Substituting $p^a=0.49$, $q^a=0.51$, and $C=0$ (starting capital) gives 
\begin{align}
z_0^A  \approx 0.47
\end{align}
Since $z_0^A < 0.5$, game A is losing because starting with $C=0$, the player is more likely to reach $C=-3$ than $C=3$.

\vskip 0.3cm
\noindent \textbf{Game B}   \\ \normalsize

Game B uses two biased coins depending on the player's current capital:

\begin{itemize}
    \item If capital \textbf{is} a multiple of 3 (we play with the unfavorable B coin):
        \begin{itemize}
            \item Head (gain $C=1$) with probability $p_{u}^{b} = 0.09$
            \item Tail (lose $C=1$) with probability $q_{u}^{b} = 0.91$
        \end{itemize}
    \item If capital is \textbf{not} a multiple of 3 (we play with the favorable B coin):
        \begin{itemize}
            \item Head (gain $C=1$) with probability $p_{f}^{b} = 0.74$
            \item Tail (lose $C=1$) with probability $q_{f}^{b} = 0.26$
        \end{itemize}
\end{itemize}

 The system of equations now becomes:
\begin{align}
z_{-3} &= 0, \quad z_3 = 1 \label{eq:b1} \\
z_{-2} &= q_{f}^{b} z_{-3} + p_{f}^{b} z_{-1} \label{eq:b2} \\
z_{-1} &= q_{f}^{b} z_{-2} + p_{f}^{b} z_0 \label{eq:b3} \\
z_0 &= q_{u}^{b} z_{-1} + p_{u}^{b} z_1 \label{eq:b4} \\
z_1 &= q_{f}^{b} z_0 + p_{f}^{b} z_2 \label{eq:b5} \\
z_2 &= q_{f}^{b} z_1 + p_{f}^{b} z_3 \label{eq:b6}
\end{align}

Solving for $z_0$:

\begin{align}
z_0 = \frac{p_{u}^{b} (p_{f}^{b})^2}{q_{u}^{b} (q_{f}^{b})^2 + p_{u}^{b} (p_{f}^{b})^2} \label{eq:z0}
\end{align}

Substituting $p_{f}^{b} = 0.74$, $q_{f}^{b} = 0.26$, $p_{u}^{b} = 0.09$, $q_{u}^{b} = 0.91$:

\begin{align}
z_0^B = \frac{(0.09)(0.74)^2}{(0.91)(0.26)^2 + (0.09)(0.74)^2} \approx 0.44\label{eq:z0val}
\end{align}

Since $z_0^B < 0.5$, game B is also losing because the probability of success is less than 50\%.

\vskip 0.3cm
\noindent \textbf{Game C: switching A and B}   \\ \normalsize

Game C is defined as follows:
\begin{enumerate}
    \item Flip a fair coin.
    \item If heads, play Game A; if tails, play Game B.
\end{enumerate}

The effective winning probabilities for game C are:
\begin{align}
p_{f}^{c} &= 0.5  p^{a} + 0.5  p_{f}^{b} = 0.5  0.49 + 0.5  0.74 = 0.615 \label{eq:pf} \\
p_{u}^{c} &= 0.5  p^{a} + 0.5  p_{u}^{b} = 0.5  0.49 + 0.5  0.09 = 0.29 \label{eq:pu}
\end{align}

The corresponding losing probabilities are:
\begin{align}
q_{f}^{c} &= 1 - p_{f}^{c} = 0.385, \quad q_{u}^{c} = 1 - p_{u}^{c} = 0.71 \label{eq:qfqu}
\end{align}

The probability of winning $C=3$ before losing $C=-3$ from state $C=0$ is:
\begin{align}
z_0^C = \frac{p_{u}^{c} (p_{f}^{c})^2}{q_{u}^{c} (q_{f}^{c})^2 + p_{u}^{c} (p_{f}^{c})^2} \label{eq:zc}
\end{align}
Substituting the values from \eqref{eq:pf}, \eqref{eq:pu}, and \eqref{eq:qfqu}:
\begin{align}
z_0^C = \frac{(0.29)(0.615)^2}{(0.71)(0.385)^2 + (0.29)(0.615)^2} \approx 0.51  \label{eq:zcval}
\end{align}

Thus, Game C is winning, $z_0^C>0.5$, because the randomized switching between Games A and B smooths out the unfavorable states of Game B while maintaining the moderate winning probabilities of Game A, resulting in an overall positive probability (greater than 50\%) of reaching $C=3$ before $C=-3$. This demonstrates the PP: combining two losing games ($z_0^A<0.5$, $z_0^B<0.5$) can create a winning strategy ($z_0^C>0.5$).

\subsection{Discrete dynamical systems}
 
For readers less familiar with the theory of nonlinear dynamics, we introduced in this subsection the fundamental elements of dynamical systems from which chaotic Parrondian models arise.

We consider a metric space $(X,d)$ and the set $C(X)$ of continuous maps $%
f:X\rightarrow X$. Given a sequence of maps $f_{n}\in C(X)$, we construct
the difference equation
\begin{align}
x_{n+1}=f_{n}(x_{n}).  \label{notaut1}
\end{align}%
For an initial condition $x_{0}\in X$, the orbit under the non-autonomous
difference Eq.~(\ref{notaut1}) is given by the recurrence%
\begin{equation*}
\left\{
\begin{array}{l}
x_{0}\in X, \\
x_{n+1}=f_{n}(x_{n}).%
\end{array}%
\right.
\end{equation*}%
We denote this orbit by $\mathrm{Orb}(x,f_{1,\infty })$, where $f_{1,\infty
}=(f_{1},f_{2},...)$. The set of limit points of the orbit is the $\omega $-limit set, denoted $\omega (x,f_{1,\infty })$. This framework of
non-autonomous discrete systems was introduced by Kolyada and Snoha in~\cite{kolsno}. We are interested in working with non-autonomous systems with the property that the number of maps in the sequence $f_{1,\infty }$ is finite, namely, $f_{n}\in \{f_{1},...,f_{k}\}\subset C(X)$. We will also assume that each map $f_{i}\in \{f_{1},...,f_{k}\}$ appears infinitely many times in the sequence $f_{1,\infty }$. Note that when the sequence $f_{1,\infty }$ is constant, that is, it consists of a single map $f$, then we have an autonomous discrete dynamical system, and we denote the orbit and limit set by $\mathrm{Orb}(x,f)$ and $\omega (x,f)$.

Non-autonomous difference equations can be found in a class of maps called
triangular~\cite{kolsno} or skew-product maps~\cite{krengel}.
Assume that $X=X_{1}\times X_{2}$, where $X_{i}$ are metric spaces for $i=1,2
$, and let $T(x,y)=(f(x),g(x,y))$, where we will assume that both maps $f\in
C(X_{1})$ and $g:X_{1}\times X_{2}\rightarrow X_{2}$ are continuous.
Writing the difference equation $(x_{n+1},y_{n+1})=T(x_{n},y_{n})$ as $%
x_{n+1}=f(x_{n})$ for the first coordinate and
\begin{equation*}
y_{n+1}=g(x_{n},y_{n}):=g_{x_{n}}(y_{n}).
\end{equation*}%
Thus, the sequence of the second coordinate can be seen as the
non-autonomous difference equation given by the initial condition $x_{0}$ dependent sequence
$(g_{x_{0}},g_{x_{1}},...)$, as discussed at large in~\cite{BlockCoppel} or~\cite{devaney}.

In respect of the initial condition it can be classified as follows: when $f(x_{0})=x_{0}$, then $x_{0}\in X$ is said to be fixed for $f\in C(X)$; alternatively, the point $x_{0}$ is periodic if there is $n\in \mathbb{N}$ such that $f^{n}(x_{0})=x_{0}$, where $f^{1}=f$ and for $n>1$, $f^{n}=f\circ f^{n-1}$. 

\begin{figure}[htbp]
\begin{center}

(a) \includegraphics[width=0.4\textwidth]{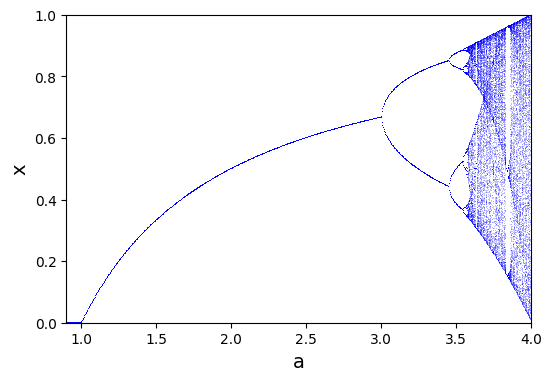} \\
(b) \includegraphics[width=0.4\textwidth]{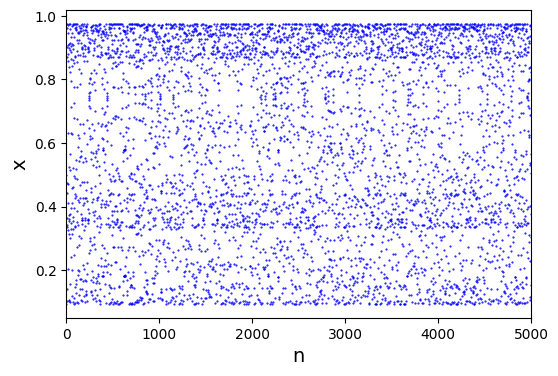}
\caption{(a) Bifurcation diagram of the logistic map. For the initial condition $x_0=0.5$, we compute orbits of length 10000 and depict the last 200 points of each orbit for $a\in [0,4]$ with step size 0.004. (b) Time series of the orbit of the logistic map with $a=3.9$.  }
\label{log1}
\end{center}
\end{figure}

Periodic orbits are the most regular orbits that a discrete dynamical system may generate. Thus, we can state that a regular dynamics is one given by a map $f\in C(X)$ such that for all $x\in X$ the set $\omega (x,f)$ is a periodic orbit.
This is the case of the family of maps $f_{a}:I\rightarrow I$, $I=[0,1]$, given by $f(x)=ax(1-x)$, $a\in (0,3]$. It is well-known~\cite{may} that if $a\leq 1$ all the orbits converge to the fixed point $x_{0}=0$, while if $a\in (1,3]$, two possible fixed points exist, $x_{0}=0$ and $x_{1}=\frac{a-1}{a}$, which are the unique $\omega $-limit sets of the associated dynamical system. Increasing the parameter $a$ produces new periodic orbits of higher
periods $2^{n}$, $n\in \mathbb{N}$, until we reach the limit value $a_{\infty }=3,5699...$ as show in the famous bifurcation diagram of Figure~\ref{log1}.

\begin{figure}[htbp]
\begin{center}
\includegraphics[width=0.4\textwidth]{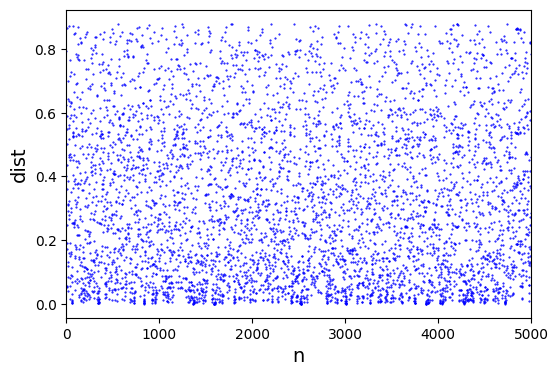}
\caption{Time series of the absolute value of the difference of two orbits with initial conditions $x_0=0.5$ and $y_0=0.1$ of the logistic map with parameter $a=3.9$.  }
\label{log2}
\end{center}
\end{figure}

Defining chaos rigorously is not straightforward, as numerous non-equivalent definitions have been introduced in the literature~\cite{devaney,Liyorke}. Herein, we will consider three different
approaches that are enough for our purposes. 
Assuming that $X$ is compact, the seminal definition of chaos by Li and Yorke~\cite{Liyorke} is based on the distance between two orbits, as Fig.~\ref{log2} shows; i.e., a subset $S\subset X$ is said to be scrambled if for any $x,y\in S$, $x\neq y$, we have
\begin{equation*}
0=\liminf_{n\rightarrow \infty }d(f^{n}(x),f^{n}(y))<\limsup_{n\rightarrow
\infty }d(f^{n}(x),f^{n}(y)),
\end{equation*}
then, the map $f$ is chaotic if there exists an uncountable scrambled set $S\subset X$.

Another definition of chaos based on a single orbit~\cite{devaney} that states that $f$ is D-chaotic if there exists $x\in X$ such that $\omega (x,f)=X$ -- i.e., the set of periodic points of $f$ --- is dense in $X$ and $f$ is sensitive to initial conditions. In other words, there is $\varepsilon >0$ such that for any $x\in X$ there is an arbitrarily close $y\in X$ and $n\in \mathbb{N}$ such that $d(f^{n}(x),f^{n}(y))>\varepsilon $.

Although the seminal definition is due to Adler, Konheim and McAndrew~\cite{adkoma}, in this review we adopt an equivalent definition introduced by Bowen~\cite{bowen1}; explicitly, for a given separation tolerance $\varepsilon >0$ and an iteration length $n \in \mathbb{N}$, a set $E \subset X$ is defined as $(n, \varepsilon, f)$-separated if for any two distinct points $x, y \in E$, there must be at least one time $k \in \{0, 1, \dots, n-1\}$ such that their orbits are separated by a distance greater than $\varepsilon$, i.e., $d(f^{k}(x),f^{k}(y))>\varepsilon$.
The quantity $s(n, \varepsilon, f)$ denotes the maximum cardinality (number of elements) of any maximal $(n, \varepsilon, f)$-separated set contained within $X$. This value measures how many points can be distinguished over $n$ iterations with a resolution of $\varepsilon$.

The topological entropy of $f$, denoted $h(f)$, is then defined as the exponential growth rate of this maximum separated cardinality as the number of iterations $n$ approaches infinity, and the separation resolution $\varepsilon$ approaches zero:
$$
h\left( f\right) =\lim_{\varepsilon \rightarrow 0}\limsup_{n\rightarrow
\infty }\frac{1}{n}\log s\left( n,\varepsilon ,f\right) .
$$
Notice that the three definitions of chaos stated here are purely
topological notions.

It is known that if $h(f)>0$, then the map $f$ is LY-chaotic~\cite{blan}. We can have deeper results if we restrict to one-dimensional maps, that is, if $f\in C(I)$, with $I=[0,1]$. On one hand, the topological entropy is a useful tool to check the dynamical complexity of a map because it is strongly connected with the notion of horseshoe~\cite[page 205]{allimi}. We say that the map $f:X\rightarrow X$ has a $k$--horseshoe, $k\in
\mathbb{N}$, $k\geq 2$, if there are $k$ disjoint subintervals $J_{i}$, $i=1,...,k$, such that $J_{1}\cup ...\cup J_{k}\subseteq f(J_{i})$, $i=1,...,k$\footnote{Since Smale's work~~\cite{smale}, horseshoes have been at the core of
chaotic dynamics, describing what we could call random deterministic systems.}. Besides, if $f$ is D-chaotic, then $h(f)>0$, but there are LY-chaotic maps with zero topological entropy~\cite{smital,BlockCoppel}. Adding some regularity conditions to the map $f$ we can strengthen the
results as follows.

We say that $f\in C(I)$ is multimodal if there exist $0=c_{0}<c_{1}<...,<c_{m}=1$ such that $f$ is strictly decreasing or strictly increasing on the subinterval $(c_{i},c_{i+1})$, $i=0,1,...,m-1$. Clearly, for $i=1,...,m-1$, we have that $c_{i}$ is a local extremum of $f$, called turning point. When $m=2$, we say that $f$ is unimodal and, if it is equal to three we say that it is bimodal. If the multimodal map $f$ is $C^{3}$ without flat turning points, we have that there are not LY-chaotic maps with zero topological entropy and hence, topological entropy is a tool to decide whether a map is LY-chaotic~\cite{baljim}. The topological entropy is hard to compute in general, but for multimodal maps with at most three turning points with different forward images we can use several algorithms that can help us to characterize topological chaos in a practical way~~\cite{Canovas3}.

All the above is a key to explain a paradoxical behavior of chaotic dynamics. 
We provide further details about this topic in Appendix~\ref{app:chaos_details}.
Because of this topological nature of chaos, there are cases in which chaos cannot be definitively detected or verified solely through numerical simulations, which rely on finite precision and sampled data. A clear example is the dynamics of the logistic map $f(x)=3.83x(1-x)$, which has positive topological entropy and is therefore topologically chaotic, yet its time series exhibits a three-periodic orbit. To characterize this phenomenon in one-dimensional dynamics, we have the so-called Lyapunov exponents~\cite{osedelec}. For $x\in X$ we define the Lyapunov exponent at $x$ as%
\begin{eqnarray*}
\lambda(f,x)&=&\lim_{n\rightarrow \infty }\frac{1}{n}\log |(f^{n})^{\prime
}(f(x))| \\&=&
\lim_{n\rightarrow \infty }\frac{1}{n}\sum_{j=1}^{n}\log |f^{\prime
}(f^{j}(x))|.
\end{eqnarray*}%

\begin{figure}[htbp]
\begin{center}

(a) \includegraphics[width=0.4\textwidth]{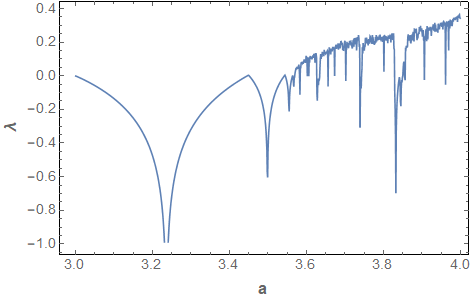} \\
(b) \includegraphics[width=0.4\textwidth]{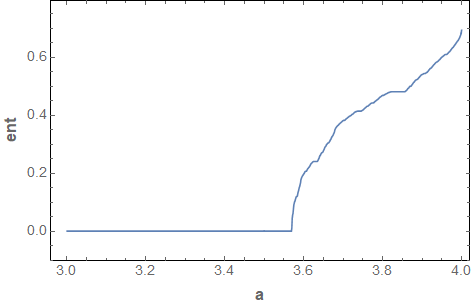}
\caption{(a) Estimation of the Lyapunov exponents of the logistic map with the initial condition $x_0=0.5$. (b) Computation of the topological entropy of the logistic map with accuracy 0.001. In both cases, we take $a\in [3,4]$ with step size 0.001. We observe for $a=3.83$ that the Lyapunov exponent is negative and the topological entropy is constant. This indicates that, although topological chaos exists, it cannot be observed in practice because almost all orbits converge to a periodic orbit. In other words, the chaotic set has Lebesgue measure equal to zero.}
\label{log3}
\end{center}
\end{figure}

Let $\{c_{1}, \dots, c_{i-1}\}$ denote the turning points of the function $f$. The Lyapunov exponent associated with a turning point $c_{j}$, denoted $\lambda (f, c_{j})$, is negative if the orbit of $c_{j}$ converges to a periodic orbit. A positive Lyapunov exponent, $\lambda(f, x) > 0$, serves as a primary numerical indicator for detecting the presence of chaotic behavior in a dynamical system. In Figure~\ref{log3}, we show the estimations of the Lyapunov exponents for the logistic map.

\section{Chaotic switching}

Chaos is undesirable in many fields of science and technology. But this is not always the case. For instance, 
chaos can be used to generate discrete sequences that can be employed to construct switching protocols as done in~\cite{bucolo2002does} where the authors have shown in standard Parrondian games the gain can be enhanced with chaotic switching. 
Later,~\cite{arena2003game} has shown that chaotic switching can also improve the capital gain in three-games as well as in $N$-games.
These works compared random and chaotic protocols under arbitrary parameter settings. 
Later,~\cite{tang2004investigation,tang2004parrondo} proposed a general fair method for comparing random and chaotic switching protocols in Parrondian games. 
In~\cite{ejlali2018parrondo,ejlali2020parrondo}, the authors considered two games played among three players with periodically and chaotically switching.

Nonetheless, it must be emphasized that nonlinear Parrondian systems, such as those we will review, are a particular class objectively enhancing chaotic switching protocols, i.e., we use the combination of (under)performances on a given observable to obtain optimal (over)performance on the same observable. Loose connections might be established with problems where we use nonlinear switching protocols to improve the results measured in another quantity; an example of this case is the work presented in~\cite{yang2025chaos,minati2025enhanced} where by applying a chaotically modulated current in a process of manganese electrodeposition the authors obtain a deposit that is smoother, more uniform, and largely free of defects -- viz., a $59\%$ reduction -- than a direct current deposit 
which implies an improvement in current efficiency with significant energy savings.

In the quantum domain,~\cite{lai2021chaotic}
showed some scenarios where chaotic switching can enhance properties of a Parrondian quantum game. They also discussed potential applications in the area of cryptography. While chaotic switching was used in discrete-time quantum walks, so far, there is no investigation of this type of switching in continuous-time quantum walks~\cite{ximenes2024parrondo,ximenes2025enhanced}

In all the cases mentioned, the following maps were used to perform chaotic switching: 
\begin{itemize}
\item 1D maps: 
Logistic, Gaussian, Tent, sinusoidal
\item 2D maps: 
Lozi, Henon    
\end{itemize}


Chaotic switching has the interesting property of being inherently aperiodic. However, there are also deterministic aperiodic protocols without chaos that have been shown to yield non-trivial results across diverse areas~\cite{macia2023alloy,pires2020quantum}.
For example, in~\cite{luck2019parrondo} the author used a switching protocol based on aperiodic sequences such as the Fibonacci sequence. Later, in~\cite{pires2024parrondo}, the authors considered other canonical aperiodic structures such as the Thue-Morse and Rudin-Shapiro sequences. 
 Aperiodic switching rules with randomness were also considered~\cite{kocarev2002lyapunov,boyarsky2005randomly}.

\section{The dynamic Parrondo's paradox }

Recall that $(X,d)$ is a metric space and assume a finite family of maps $f_{1},...,f_{k}\in C(X)$. We consider the set $\Sigma _{k}=\{(\sigma
_{n}):\sigma _{n}\in \{1,...,k\},\ n\in \mathbb{N}\cup \{0\}\}$, which is a compact metrizable topological space. Taking the product space $\Sigma _{k}\times X$, we can define a triangular map $T:\Sigma _{k}\times X\rightarrow \Sigma _{k}\times X$ such that for any $((\sigma _{n}),x)\in \Sigma _{k}\times X$, its forward image is given by%
\begin{equation*}
T((\sigma _{n}),x):=\left( \sigma (\sigma _{n}),T_{\sigma _{0}}(x)\right) ,
\end{equation*}%
where $\sigma (\sigma _{n})=(\sigma _{n+1})$ is the shift map that removes the symbol $\sigma _{0}$ from the sequence $(\sigma _{n})$ and the fiber map%
\begin{equation*}
T_{\sigma _{0}}(x):=f_{j}(x)
\end{equation*}%
provided $\sigma _{0}=j\in \{1,...,k\}$. Thus, we easily see that for any sequence $(\sigma _{n})$, we have a non-autonomous discrete system in which a finite number of maps are iterated. In addition, iterating the shift map, we can assume without loss of generality that every map in the non-autonomous system generated by $\sigma ^{m}(\sigma _{n})$, $m\in \mathbb{N}$, appears infinitely many times in the sequence of maps constructed from that symbolic sequence. We define the subset $\Sigma _{k}^{\ast }\subset \Sigma _{k}$ of sequence with the property that for any $j\in \{1,...,k\}$ the cardinality $\#(\{j\}\cap \{\sigma _{n}:n\in \mathbb{N}\})$ is either zero or infinite. Note that $\sigma \left( \Sigma _{k}^{\ast }\right) \subset \Sigma _{k}^{\ast }$, and then we can restrict our map $T$ to the set $\Sigma _{k}^{\ast }\times X$. 
This restricted map is denoted by $T^{\ast }$. This is the general framework in which the dynamic Parrondo paradox takes place.

This dynamic Parrondo paradox draws on the game-theory paradox. In general, let us assume that the maps $f_{1},...,f_{k}$ are dynamically simple (resp. complex). Then, we have paradox for $(\sigma _{n})\in \Sigma _{k}^{\ast }$ if the non-autonomous system generated by $(\sigma _{n})$ is dynamically complicated (resp. simple). Later, we will specify  what simple or complex means for a non-autonomous discrete system, but in general it will be clear when the sequence $(\sigma _{n})$ is periodic, which is the initial setting of the paradox~~\cite{almeida2005can}.

The most common formulation of the dynamic Parrondo paradox involves two maps, $f$ and $g$. The paradox occurs when the dynamics of $f$ and $g$ are simple (resp. chaotic) and the dynamics of the periodic sequence $(f,g,f,g,...)$ are chaotic (resp. simple). First, note that if we consider the set $S=\{s_{1},s_{2}\}\subset \Sigma _{2}^{\ast }$, where $s_{1}=(1,2,1,2,...)=\sigma ^{2}(s_{1})$ and $s_{2}=(2,1,2,1,...)=\sigma (s_{1})$, we can consider the map $T_{2}=T|_{S\times X}$. This statement of the paradox fits within the general framework. On the other hand, note that if $x\in X$, its orbit by the $(f,g,f,g,...)$ is given by the sequence%
\begin{equation*}
(x,f(x),(g\circ f)(x),(f\circ g)(f(x)),(g\circ f)^{2}(x),(f\circ
g)^{2}(f(x)),...).
\end{equation*}%
Then, its dynamical behavior can be deduced from that of the compositions $f\circ g$ and $g\circ f$. Several authors studied how to relate the dynamics of both compositions~\cite{kolsno} and showed that for the most popular dynamical properties (e.g., topological entropy, LY-chaos, periodicity,...) the two compositions are equivalent. For instance, $h(g\circ f)=h(f\circ g)$, or the map $g\circ f$ is LY-chaotic if and only if $f\circ g$ is so. There are some exceptions to this rule (for example, when topological sequence entropy is concerned~\cite{bcj1999}), but in this paper, we can consider the dynamics of both compositions as equivalent. This implies that we can consider $(f,g,f,g,...)$ or $(g,f,g,f,...)$ to state the paradox and that it can be stated by replacing $(f,g,f,g,...)$ with $g\circ f$ (or $g\circ f$).  Note that it is easy to extend this paradox to more than two maps.

\subsection{Local  Parrondo's Paradox  and fixed point dynamics}

Here we consider smooth enough maps $f_{j}:A\rightarrow A$, where $%
A\subseteq \mathbb{R}^{n}$, $j=1,...,k$. 
Assume all maps share a common fixed point 
$x_{0}\in A$ such that $f_{j}(x_{0})=x_{0}$, $j=1,...,k$. First, we introduce some specific notation for this case~\cite{Elaydi0}.

A fixed point $x_{0}$ of a map $f:A\rightarrow A$ is said to be locally asymptotically stable (LAS) if stable and locally attractive, that is, for any $\varepsilon >0$ there is $\delta >0$ such that if $||x-x_{0}||<\delta $ then $||f^{n}(x)-x_{0}||<\varepsilon $ for all $n\geq 1$ (stability) and there is $\eta >0$ such that if $||x-x_{0}||<\eta $, then $\lim_{n\rightarrow \infty }f^{n}(x)=x_{0}$ (attractive). The fixed point is globally asymptotically stable (GAS) if it is LAS and $\lim_{n\rightarrow \infty }f^{n}(x)=x_{0}$ for all $x\in A$. The fixed point is a repeller if there exist $\varepsilon _{0}>0$ such that for all $\varepsilon \in (0,\varepsilon _{0})$ and all $x\neq x_{0}$ such that $||x-x_{0}||<\varepsilon $ there exists $n=n(x_{0})$ such that $||f^{n}(x)-x_{0}||>\varepsilon $.

Within the context of dynamical systems, a `losing' strategy can also be an attracting fixed point, whereas a `winning' one is a repelling fixed point. With that idea in mind, the authors of~\cite{cat1} consider that the  PP  is happens when the common fixed point is LAS (resp. repeller) for $f_{j}$, $j=1,...,k$ and it is a repeller (resp. LAS) for $f_{k}\circ ...\circ f_{1}$. As pointed out in~\cite{cat1}, when the fixed points are hyperbolic, the paradox is not possible, but the possibility remains open when they are non-hyperbolic. Recall that hyperbolicity implies that there are not eigenvalues with modulus one in the Jacobian matrix at the fixed point. The following results from~\cite{cat1} show that  PP  can happen in this context.

\begin{theorem}
\label{las1}The following statements hold:

\begin{enumerate}
\item If $f_{1}$,f$_{2}:A\subseteq \mathbb{R}\rightarrow A$ are analytical and share a common fixed point $x_{0}$ which is LAS (resp. repeller). Then $x_{0}$ is either LAS (resp. repeller) or the composition $f_{2}\circ f_{1}$ is semi-stable (i.e., the LAS condition only happens in left side or right side of the interval $(x_{0}-\varepsilon ,x_{0}+\varepsilon )$, $\varepsilon >0$, but not in both sides). Moreover, both possibilities happen.

\item For $k\geq 3$, there are $k$ polynomial maps $f_{j}:A\subseteq \mathbb{R}\rightarrow A$, $j=1,...,k$, with a common fixed point which is LAS (resp. repeller) and it is repeller (resp. LAS) for $f_{k}\circ ...\circ f_{1}$.
\end{enumerate}
\end{theorem}

The above result indicates that the number of maps is crucial, as the paradox appears more likely to occur as the number of iterated maps increases. Note that it is given in the setting of one-dimensional maps, which is the more common context in which the paradox has been studied. The next result shows that the dimension of the phase space also has influence on the paradox.

\begin{theorem}
\label{las2}There exist polynomial maps $f_{1}$ and $f_{2}$ in $\mathbb{R}^{2}$ sharing a common fixed point x$_{0}$ which is a LAS (resp. a repeller) fixed point for both of them, and such that p is repeller (resp. LAS) for the composition map $f_{2}\circ f_{1}$.
\end{theorem}

Again, as shown in~\cite{cat1}, the result of Theorem \ref{las2} is possible when the number of maps in the iteration is increased. We also refer the readers to the references~\cite{cat3}, where this problem is analyzed for continuous time models.

On the other hand, the authors of~\cite{cat2} considered that a pair of maps $f, g \in \mathcal{C}$ are said to exhibit a dynamical  PP  if they have a common fixed point that is asymptotically stable for both $f$ and $g$ but is a repeller for the composite maps $g\circ f$ and $f\circ g$. They do so, asserting that for any $k\ge2$, there are pairs of homeomorphisms from $\mathbb{R}^{k}$ to itself that exhibit the dynamical  PP, but that for $k=1$ the paradox never arises.

\subsection{Topological Parrondo's Paradox}

The notions of chaos and dynamical simplicity can be seen from the lens of topology.
Hence, the standard  PP  for two maps $f,g:X\rightarrow X$ has also a topological nature. Since it has been studied mainly for
one-dimensional maps, we assume that $X\subset \mathbb{R}$ is a closed
interval of the real line. A first approach to the paradox allows us to
state that for most of the topological dynamics notions this paradox happens. This fact was proved in~\cite{canovas2006dynamic} and, for instance, in the case of the topological entropy, it can be stated as follows: there exist continuous interval maps $f$ and $g$ such that $h(f)=h(g)=0$ (resp. $\min \{h(f),h(g)\}>0$) and $h(f\circ g)>0$ (resp. $h(f\circ g)=0$).

This result was proved by taking a partition of the interval $X$ and defining the maps on this partition with the required dynamical properties so that the composition can show a completely different dynamical behavior. If we do not allow that our maps will be defined in this way, that is, we consider a parametric family of interval maps $f_{a}:[0,1]\rightarrow \lbrack 0,1]$, where $a$ is the parameter, and we ask for a similar question: are there two parameter values $a,b$ such that $h(f_{a})=h(f_{b})=0$ (resp. $\min \{h(f_{a}),h(f_{b})\}>0$) and $h(f_{b}\circ f_{a})>0$ (resp. $h(f_{b}\circ f_{a})=0$), then we realized that answering this question is not that simple.

\begin{figure}[htbp]
\begin{center}

(a) \includegraphics[width=0.4\textwidth]{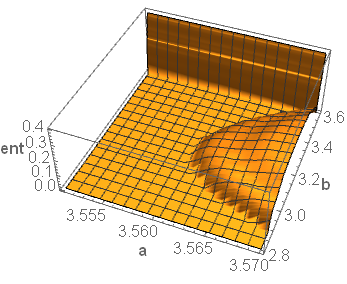} \\
(b) \includegraphics[width=0.4\textwidth]{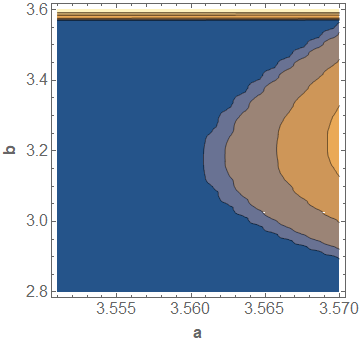}
\caption{(a) Topological entropy of the map $f_{a}\circ f_{b}$ with accuracy $10^{-4}$ for $a\in [3.55,3.57]$ and $b\in [2.8,3.6]$. (b) Level curves of the computations. As $h(f_a)=0$ for $a<3.5699...$, the existence of positive entropy inside the rectangle $[3.55,3.57]\times [2.8,3.6]$ proves the existence of the Parrondo paradox "simple+simple$\to$chaos".}
\label{entropia}
\end{center}
\end{figure}

In this present form this question was studied first in~\cite{canovas2013revisiting}. Here, the logistic family $f_{a}(x)=ax(1-x)$, $x\in \lbrack 0,1]$ and $a\in (0,4]$ was considered to prove that there is a set $P\subset (0,4]^{2}$ such that for any $(a,b)\in P$, we have that $h(f_{a})=h(f_{b})=0$ while $h(f_{b}\circ f_{a})>0$. The shape of this set can be seen in Figure~\ref{entropia}.

It is unclear whether the reverse paradox is possible, that is, if there are $(a,b)\in (0,4]^{2}$ such that $\min \{h(f_{a}),h(f_{b})\}>0$ and $%
h(f_{b}\circ f_{a})=0$. Note that the logistic map is $C^{\infty }$ with
non-flat turning point, and hence positive topological entropy and LY-chaos are equivalent notions for the family. In~\cite{canovas2018ricker}, a similar result is obtained for the Ricker model $g_{r}(x)=xe^{r-x}$, $x\geq 0$, $r>0$.

For the family of tent maps $t_{s}$, $s\in (0,2]$, defined by%
\begin{equation*}
t_{s}(x):=\left\{
\begin{array}{ccc}
sx & \mathrm{if} & x\in \lbrack 0,1/2], \\
s-sx & \mathrm{if} & x\in (1/2,1],%
\end{array}%
\right.
\end{equation*}%
the situation is completely different. It is know that $h(t_{s})=\max
\{0,\log s\}$~\cite[Chapter 4]{allimi} and it was proved in
~\cite{canovas2021parrondo} that for $s_{1},s_{2}\in (0,2]$ we have that $h(t_{s_{1}}\circ t_{s_{2}})=\max \{0,\log \left( s_{1}s_{2}\right) \}$. Thus, if $\max \{s_{1},s_{2}\}\leq 1$, then $h(t_{s_{1}})=h(t_{s_{2}})=h(t_{s_{1}}\circ t_{s_{2}})=0$, while if $\min \{s_{1},s_{2}\}>1$, then we have that $\min 
\{h(t_{s_{1}}),h(t_{s_{2}}),h(t_{s_{1}}\circ t_{s_{2}})\}>0$, and therefore no paradox is possible for this family of maps. The tent map has a constant slope $s$. If we extend this family to maps with two possible slopes in absolute value, namely,%
\begin{equation*}
t_{s,p}(x):=\left\{
\begin{array}{ccc}
sx & \mathrm{if} & x\in \lbrack 0,p], \\
\frac{1-sp}{1-p}x+p\frac{s-1}{1-p} & \mathrm{if} & x\in (p,1],%
\end{array}%
\right.
\end{equation*}%
where $p\in (0,1)$ and $s\in (0,1/p)$, then it was showed in~\cite{canovas2021parrondo} that there exists a subset in the parameter space $(s_{1},s_{2},p)$ such that $ h(t_{s_{1},p})=h(t_{s_{2},p})=0$ and $h(t_{s_{1},p}\circ t_{s_{2},p})>0$.
Again, it is unclear whether the converse paradox "chaos+chaos$\to$simple" is
possible for this family.

\begin{remark}
All the above results of type "simple+simple$\to$chaos" have been obtained using numerical algorithms that allow for computing the topological entropy with a prescribed accuracy~\cite{Canovas3} for a survey on this topic. Hence, we consider zero entropy maps and show that the topological entropy of their composition exceeds the prescribed accuracy. Proving the converse paradox, if possible, is more complicated since no numerical algorithm can assure that the topological entropy is zero.
\end{remark}

\begin{remark}
Under the assumption of commutativity, that is, $f\circ g=g\circ f$, it was proved in~\cite{cali} that is both maps are piecewise monotone, then $h(f\circ g)\leq h(f)+h(g)$. Thus, if $h(f)=h(g)=0$, then $h(f\circ g)$ must be also zero and no paradox is possible. This argument was expanded in~\cite{canovas2010analyzing} to show that sequences of piecewise monotone commuting maps $f_{1},...,f_{k}$ with zero topological entropy cannot display positive topological entropy.
\end{remark}

Finally, we finish this section by considering the paradox for the logistic map when more than two maps are iterated. Namely, in~\cite{canovas2016computing} it was shown that if $a,b,c\in (0,4]$ such that $h(f_{a})=h(f_{b})=h(f_{c})=0$, then the topological entropy of $f_{c}\circ f_{b}\circ f_{a}$ can be positive.
Moreover, it can be positive for parameter values that do not display
positive entropy for $f_{b}\circ f_{a}$. So  it seems that the paradox is easily exhibited if the number of maps increases.

\subsection{Observable Parrondo's Paradox}

This section is devoted to analyze the physical observability of the  PP. Roughly speaking, it is the one we observe when we make numerical simulations, and for this reason, it appears in the seminal first works on this topic~\cite{kocarev2002lyapunov,almeida2005can}).

To fix ideas, we consider two smooth enough interval maps $f,g:X\subset \mathbb{R}\rightarrow X$. To simplify, we will assume they are piecewise monotone and $C^{3}$ with non-flat turning points. Then, the attractors of the maps $f$, $g$ and $f\circ g$ can be a periodic orbit, or periodic mixing subintervals or Cantor type sets, but the last one is quite difficult to observe in a numerical simulation. So, typically we see periodic orbits (simple dynamics) or periodic intervals (complex dynamics). If a map has zero topological entropy, only periodic orbits can be observed, but positive topological entropy can produce both kinds of attractors: if the Lyapunov exponent is negative, we will have periodic orbits, while positive Lyapunov exponents are taken as a proof of complex dynamics, produced by the periodic mixing of subintervals.

To the best of our knowledge, the first work to explicitly state the dynamic PP was~\cite{kocarev2002lyapunov}. They considered two
piecewise linear maps, namely%
\begin{align}
f_{1}(x) &=\left\{
\begin{array}{lll}
\frac{1}{a}x & \mathrm{if} & x\in \lbrack 0,a), \\
\frac{a}{1-a}x-\frac{a^{2}}{1-a} & \mathrm{if} & x\in (a,1],%
\end{array}%
\right. 
\\
f_{2}(x)&=\left\{
\begin{array}{lll}
\frac{a}{1-a}x+1-a & \mathrm{if} & x\in \lbrack 0,1-a), \\
\frac{1}{a}x-\frac{1-a}{a} & \mathrm{if} & x\in (1-a,1].%
\end{array}%
\right.
\end{align}%
Although these maps are not continuous, their dynamics work as continuous ones in terms of chaos, so it makes sense to study their dynamics using the previous background. Then, fixing a non-negative $p\leq 1$, they establish the iteration of the map $f_{1}$ with probability $p$ and of $f_{2} $ with probability $1-p$. The Lyapunov exponent of both maps is computed and is equal to%
\begin{equation*}
\lambda(f_i,x)=\frac{1}{2-a}\log \left( \frac{1}{a}\right) +\frac{1-a}{2-a}\log \left(
\frac{a}{1-a}\right) ,
\end{equation*}%
with $i=1,2$. They showed that the Lyapunov exponents can be positive while the iteration process defined above can display a negative Lyapunov exponent for $a=0.2$ and $p=0.5$. Then, they established the dynamical PP `chaos+chaos$\to$simple'.

Next, in~\cite{almeida2005can} the authors consider the real quadratic maps $f_{i}(x)=x^{2}+c_{i}$ with initial conditions inside the invariant intervals
$\left[ -\frac{1+\sqrt{1-4c_{i}}}{2},\frac{1+\sqrt{1-4c_{i}}}{2}\right] $
for suitable values of $c_{i}\leq 1/4$. Then, they consider the periodic
sequence $(f_{1},f_{2},f_{1},f_{2},...)$ and combining Lyapunov exponents
and the existence of Misiurewicz points, they also show the existence of the `chaos+chaos$\to$simple' paradox. Misiurewicz points are parameter values for which the orbit of the turning point is eventually periodic and, if they are unstable then the Lyapunov exponent of the turning point will be positive indicating a chaotic behavior. Thus, they found examples of two maps with Misiurewicz points for which the periodic sequence exhibits simple dynamics.

It is worth noting that the quadratic family and the logistic family are conjugate and have the same dynamics; in particular, their topological entropies coincide. The examples in~\cite{almeida2005can} are located in the parameter region where the topological entropy is positive, that is, $h(f_{1}\circ f_{2})>0$ and hence chaotic, although its chaotic set is located in a set of zero Lebesgue measure. Thus, in~\cite{canovas2013revisiting} the authors extend this idea by considering both the topological entropy and the Lyapunov exponents and found examples of the paradox "simple+simple$\to$chaos" for both zero and positive topological entropy logistic maps, as well as, "complex+complex=simple" for positive entropy maps, where simple means a negative Lyapunov exponent of the turning point. The question remains open as to whether it is possible to obtain a zero topological entropy map from two maps with positive topological entropy.

The switching of logistic maps has been analyzed in several papers by Peacock-L\'{o}pez and collaborators~~\cite{maier2010switching,peacock2011seasonality,silva2017seasonality}, providing a large number of examples addressing the paradox. These papers focus on the paradox for observable chaos. In addition, they also provide examples of the paradox for modified logistic maps~\cite{levinsohn2012switching} and several population dynamics models~\cite{mendoza2018switching}. Similar examples of the paradox for modified logistic maps can also be found in~\cite{yadav2015modified,agarwal2020chaotic,rani2016parrondo}.

It is remarkable that in~\cite{peacock2011seasonality,silva2017seasonality} it is analyzed the existence of the paradox for periodic sequence of maps of period greater than two, in particular four and twelve periods. The question of whether the number of iterated maps influences the paradox was analyzed in~\cite{canovas2017periodic}, where the existence of the paradox "simple+simple$\to$chaos" was established for periodic sequences formed with two logistic maps with different period sizes. It was found that not only the period size but also the order in which the maps are iterated influences the existence of the paradox.

The reader may find this section with plenty of one-dimensional examples but not much in higher dimensions. There are examples of the paradox -- e.g.~\cite{mendoza2018parrondian} -- in which the authors show evidence of the paradox for two-dimensional models coming from ecology, from numerical simulations. The fact that the dynamics of two-dimensional dynamics is not really well known explains this lack of examples, which may deserve further investigation.

\section{Interdisciplinary connections}

Beyond the theoretical frameworks of physics and game theory,  PP  has found practical applications that offer a powerful conceptual tool for understanding complex systems in various disciplines.

\subsection{Physical and Mathematical}

\textbf{Parameter switching algorithm. ---} 
Danca et al introduced the following system of differential equations in~\cite{danca2013convergence}
\begin{equation*}
x^{\prime }(t)=f(x(t))+pAx(t),
\end{equation*}%
where $f:\mathbb{R}^{n}\rightarrow \mathbb{R}^{n}$ is a nonlinear (at
least) continuous vector-valued function, $A$ is a non-zero square matrix of
size $n$ with real entries, and $p$ is a real control (bifurcation)
parameter. If we consider $p_{1},...,p_{k}$ different values of the
parameter and $m_{1},...,m_{k}>0$ so that the parameter $p_{1}$ acts for $%
t\in \lbrack 0,m_{1})$, then $p_{2}$ for $t\in \lbrack m_{1},m_{1}+m_{2})$
and inductively $p_{k}$ for $t\in \lbrack m_{1}+...+m_{k-1},m_{1}+...+m_{k})$
and repeat this scheme periodically for $t>m_{1}+...+m_{k}$, then it is proven that the resulting system behaves like the autonomous one with parameter
\begin{equation*}
\widetilde{p}=\frac{\sum_{i=1}^{k}m_{i}p_{i}}{\sum_{i=1}^{k}m_{i}}.
\end{equation*}
Using this fact, the  CCO and OOC paradoxes for the Lorenz system can be written as
\begin{equation*}
\left\{
\begin{array}{l}
x_{1}^{\prime }=10(x_{1}-x_{1}), \\
x_{2}^{\prime }=-x_{1}x_{3}-x_{2}+px_{1}, \\
x_{3}^{\prime }=x_{1}x_{2}-\frac{8}{3}x_{3}.%
\end{array}%
\right.
\end{equation*}

These ideas have been adapted to discrete systems in~\cite{danca2014generalized} and to
fractional differential equations in~\cite{danca2016,tang2016emulating}.

\textbf{Chaos control. --- } 
In nonlinear  dynamics, the principle behind  PP  has been extended to construct the `most attracting curve' on the parameter space, where a transition from simple to chaotic dynamics takes place, allowing chaotic control. In Ref.~\cite{gupta2022}, the authors apply this concept to show that for a certain range of deformation parameters, $\epsilon > 0$, the phase transition to chaos occurs earlier than in canonical Hénon-like maps, which explains the paradoxical behavior. 

Explicitly, they deformed the canonical Hénon-like map,
\begin{align}
x_{n+1} &= f_{a}(x_{n}) - b\,y_{n} \\
y_{n+1} &= x_{n}
\end{align}
where $a \ge 1$, $b > 0$, and $f_{a}(x)=a-x^{2}$ applying a $q$-deformation inspired by the Tsallis statistics algebra~\cite{borges2021along} to the state variable $x_n$, defined as
\begin{eqnarray}
    g_{q}(x) &=& [x]_{q} = \frac{x}{1+(1-q)(1-x)}, 
    \label{eq:q-deformed} \\
    && \nonumber \\
    \left[ x \right] _{\epsilon}&=&\frac{x}{1+\epsilon(1-x)},
\end{eqnarray}
with $1-q=\epsilon$ . This yields the $q$-Hénon map,
\begin{align}
\mathcal{H}_{a,b,\epsilon} \left(\begin{matrix}x\\ y\end{matrix}\right)=\left(\begin{matrix}F_{a,\epsilon}(x)-b\,y\\ \frac{x}{1+\epsilon-\epsilon x}.\end{matrix}\right),
\end{align}
where $F_{a,\epsilon}(x)=a-\frac{x^{2}}{(1+\epsilon-\epsilon x)^{2}}$.

The fixed points of this map are given by the roots of the cubic polynomial,
\begin{align}
\mathcal P (y) = \epsilon y^{3}+(1+b\epsilon)y^{2}+(1+b+\epsilon-a\epsilon)y-a,
\end{align}
for the $y$-coordinate.
For $\epsilon=0$, the system has two roots, corresponding to the fixed points of the canonical Hénon map, while for $\epsilon \neq 0$, the polynomial has three roots for $y$, which can be real or imaginary. This means that as the parameter $\epsilon$ varies, the fixed points can bifurcate and disappear, leaving only one fixed point for $\epsilon > \epsilon_{*}$. With that, it is possible to construct superstable periodic points of period $2^{n}$ for the q-Hénon map with the critical point of the map located at $x=0$. The sequence of superstable points $\{a_{0}^{2^{n}}\}$ is found by solving $F_{a,\epsilon}^{2^{n}}(0)=0$, which converges to a value $a_{0}^{*}$ where the degenerate q-Hénon map undergoes a doubling period accumulation doubling. This is then extended to the case of the q-Hénon map considering variations of the parameter $b$ and using the Newton algorithm to find the periodic points that satisfy $\mathcal{H}^{2n}\left(\begin{matrix}x\\ y\end{matrix}\right)-\left(\begin{matrix}x\\ y\end{matrix}\right)=0$ and the superstable condition that the trace of the Jacobian is zero, $\Tr J(\mathcal{H}^{2n})=0$.

Noor's logistic map was also studied in relation to Parrondo's paradox~\cite{yadav2016parrondo}. The alternate iteration of two chaotic logistic maps for different values of $\alpha$, $\beta$, and $\gamma$ resulted in an expansion of the stability domain. Although chaotic values were still observed, the application of the paradox controlled the chaotic dynamics of the system.

In~\cite{shi2015chaos}, the authors established fundamental relationships between chaotic behaviors in discrete periodic systems and their autonomous induced systems, demonstrating equivalence under certain conditions to Li-Yorke, Wiggins, and Devaney chaos. It provides criteria for chaos and sufficient conditions for the absence of chaos in periodic systems, showing that finite-dimensional linear periodic systems are not chaotic in any Li-Yorke or Wiggins sense. Furthermore, the study explores the surprising phenomenon of "Order + Order = Chaos", presenting examples where non-chaotic systems combine to produce chaotic behavior, including cases where globally asymptotically stable maps result in chaos. The work emphasizes the dependence of chaotic dynamics on both the properties of individual maps and their compositional order, raising open questions about the invariance of chaotic behaviors under different map orders.

\textbf{Chaos anticontrol. --- } In~\cite{romera2007deterministic}, two different ways of creating chaos in systems that operate in discrete steps are presented, giving shape to the anticontrol of chaos through the phenomenon 
`Order + Order $\to$ Chaos'. More specifically, in the Random Anti-control Algorithm, the control parameter $p$ of a discrete dynamic system (such as the logistic map) is randomly (or stochastically) alternated between two or more values. Each of these individual values leads the system to exhibit a stable (non-chaotic) dynamic. The random combination of these periodic dynamics leads to a chaotic dynamic. In the Deterministic Anti-control Algorithm, the control parameter $p$ is deterministically (periodically) alternated between two values. The study tests these methods on the logistic map, showing that it is possible to generate chaos even starting from periodic movements.

\textbf{$q$-deformations. --- }
The Parrondo paradox `simple+simple=complex' appears in a natural way in the so-called deformations of discrete dynamical systems. In the seminal paper~\cite{jagannathan2005,banerjee2011}, the authors consider the composition of the logistic map $f_{a}(x)=ax(1-x)$ and the $q$-deformed number, $q\in (-\infty ,2)$, given by Eq.~(\ref{eq:q-deformed}).
The $q$-deformed logistic map is
then given by $f_{a}\circ g_{q}$. These authors were interested in the richer dynamics produced by this new two-parameter family of maps, which includes the coexistence of attractors. It can be noticed that the map $g_{q}$ is a homeomorphism on $[0,1]$, and hence displays zero topological entropy. In Ref.~\cite{canovas2019}, it was proved that $h(f_{a}\circ g_{q})>0$ for parameter values of $a$ for which $h(f_{a})=0$; in other words, this is another example of the topological Parrondo paradox \textit{OOC}. Moreover, in Ref.~\cite{canovas2022}, one of us (JSC) showed that $g_{q_{1}}\circ g_{q_{2}}=q_{2-(2-q_{1})(2-q_{2})}.$ Thus, the paradox can be studied by increasing the number of maps $g_{q}$.

\textbf{Fractality. --- } 
As chaos, especially the edge of chaos, is often related to fractal geometry, the Parrondian concept can be extended to fractal research. This was done by~\cite{wang2017preliminary} by combining two complex maps with disconnected Julia sets that are able to generate a connected Julia set. Analytically, they introduce an alternated complex map,
\begin{align}
P_{c_{1}, c_{2}}: z_{n+1} = \begin{cases} z_{n}^{2}+c_{1}, & \text{if } n \text{ is even} \\ z_{n}^{2}+c_{2}, & \text{if } n \text{ is odd} \end{cases}
\end{align}
where the filled Julia set $K(P_{c_{1}, c_{2}})$ is the set of initial points $z_0$ for which the orbit remains bounded and the Julia set $J(P_{c_{1}, c_{2}})$ is the boundary of the filled Julia set. The connectivity of this alternated Julia set is shown to depend on the boundedness of the critical points of an equivalent polynomial system, which has critical points at $0$ and $\pm\sqrt{-c_1}$.
Dividing the Connectivity Loci of $J(P_{c_{1},c_{2}})$ into three different states of connectivity, namely `connect', `disconnected', and `totally disconnected' when the orbits of $0$ and $\sqrt{-c_1}$ are bounded, one of the orbits is bounded and the other is unbounded, and both orbits are unbounded, respectively.

This work then aimed to graphically verify the `disconnected + disconnected = connected' phenomenon -- in alignment with the parrondian concept -- by finding two parameters $c_1$ and $c_2$ that individually produce disconnected Julia sets -- $J(P_{c_1})$ and $J(P_{c_2})$ --, but produce a connected Julia set when alternated, $J(P_{c_1,c_2})$.

Similarly, the paper also explores the opposite phenomenon, "connected + connected = disconnected," by choosing two parameters that individually produce connected Julia sets but result in a totally disconnected Julia set when alternated.

In the previous work~\cite{yadav2015alternate}, this approach was further explored to generate alternate superior-type Julia sets with different structures: connected, disconnected, or totally disconnected. The alternate superior Julia set is generated by alternating the iteration of quadratic or cubic maps in the Upper Orbit. This method yields fractal sets that are significantly larger and thicker than those obtained by the Picard iteration.

\textbf{Chaos and order in quantum systems. --- }
In the same way, chaos can be found in quantum systems, also Parrondian CCO and OOC effects can be found in the quantum realm. A first known case is that conveyed by~\cite{panda2021order},  wherein the authors study the chaotic and periodic nature of cyclic quantum walks (QW) using a discrete time QW, the state of the walker is defined by the tensor product of the position and coin Hilbert spaces, $H_P \otimes H_c$. For a qubit coin with states $|0\rangle$ and $|1\rangle$, the general unitary coin operator
\begin{align}
C_{2}(\rho,\alpha,\beta)=\begin{pmatrix}\sqrt{\rho} & \sqrt{1-\rho}\, e^{i\alpha}\\ \sqrt{1-\rho} \, e^{i\beta} & -\sqrt{\rho } \, e^{i(\alpha+\beta)}\end{pmatrix},
\label{eq:coin}
\end{align}
where $0 \le \rho \le 1$ and $0 \le \alpha, \beta \le \pi$. The full unitary operator for the quantum walk, $U_k$, is a combination of the shift operator, $S$, and the coin operator, $C_2$,
\begin{align}
U_{k}=S\cdot(I_{k}\otimes C_{2})
\label{eq:2}
\end{align}
A QW is periodic if, after $N$ steps, the walker returns to its initial state, which means $U_k^N|\psi_i\rangle=|\psi_i\rangle$ for any arbitrary initial quantum state $|\psi_i\rangle$, which equivalent to the condition that all the eigenvalues of the unitary operator $U_k$ satisfy $\lambda_{j}^{N}=1, \forall 1\le j\le 2k \quad \text{or} \quad U_{k}^{N}=I_{2k}$; however, if this condition is not met, the QW is considered chaotic.

Following a parrondian roadmap, the authors use sequences of two different unitary operators, $A = C_{2}(\rho,\alpha,\beta)$ and $B \equiv C_{2}(\rho ^\prime,\alpha ^\prime,\beta ^\prime)$; individually, each produces chaotic QWs, but applied in an alternating fashion, the results in Ref.~\cite{panda2021order}  demonstrate that combinations of the two operators can generate a periodic QW, effectively creating a \textit{CCO} phenomenon in a quantum system. The results are shown through plots of the probability of the walker returning to the origin for 3-cycle and 4-cycle graphs. This method of mixing quantum chaotic walks is indicated as a strong formalism to generate a secure encryption-decryption mechanism in quantum key generation and cryptography~\cite{di2019perfect,vlachou2015quantum}.

As already presented, chaotic switching can be superior for various optimization problems; this was also studied by~\cite{lai2021chaotic} within the context of quantum chaotic switching for which they introduced a deterministic and reversible process that is also sensitive to initial conditions in the spirit of parrondian games to quantum Parrondo's games. The Parrondo effect is demonstrated by choosing two specific coin operators, $\hat{\mathcal{C}}_{A}$ and $\hat{\mathcal{C}}_{B}$, which are Hadamard-like matrices
\begin{align}
\hat{\mathcal{C}}_{A}=\frac{1}{\sqrt{2}}\begin{pmatrix}1&1\\ 1&-1\end{pmatrix} \quad \text{and} \quad \hat{\mathcal{C}}_{B}=\frac{1}{\sqrt{2}}\begin{pmatrix}-1&i\\ -i&1\end{pmatrix}
\label{eq:4}
\end{align}
For chaotic switching, the authors use the logistic and sinusoidal, $x_{n+1}=ax_{n}^{2}sin(\pi x_{n})$,  maps to generate the sequence for choosing between operator $A$ and $B$ and from this they propose a protocol for a quantum parrondian game with chaotic switching and illustrate its use in semiclassical encryption. This work shows that  PP  can appear under a chaotic switching scheme, which offers an advantage over previously studied methods due to its deterministic, reversible, and sensitive-to-initial-conditions nature.

\subsection{Machine Learning}

\textbf{Artificial Neural Networks. --- }
In the framework of neural dynamics, the so-called integrate-and-fire mechanism~\cite{campbell1999synchrony} is a simplified but powerful model used in computational neuroscience to describe how a neuron generates an electrical impulse, or spike. 
In Ref.~\cite{takahashi2018simple}, the authors apply it to a Bifurcating Neuron~\cite{lee2001bifurcating} defined by a piecewise operation
\begin{equation}
\label{eq:physical}
\begin{cases}
C \frac{dv}{dt} = I & \text{for } v(t) < V_{T} \quad \text{(Integration)} \\
v(t_{+}) = B(t_{+}) & \text{if } v(t) = V_{T} \quad \text{(Fire and Reset)}
\end{cases}
\end{equation}
where $I > 0$ is the constant current, $V_{T}$ is the constant threshold voltage, and $B(t)$ is the periodic base signal, $B(t+T)=B(t)$. This means that the voltage of a capacitor increases at a constant rate until it hits $V_T$ at which point it resets. The reset voltage is determined by an external periodic base signal, $B(t)$, that is a superposition of two triangular signals and a DC component
\begin{equation}
\label{eq:basesignal_def}
B(t) = E_{1}(t) + E_{3}(t) - E_{0}.
\end{equation}
$E_1(t)$ has period $T$, and $E_3(t)$ has period $T/3$.
This makes the complexity of the output spike train entirely governed by the structure of this base signal.

In their paper~\cite{takahashi2018simple} the authors tested three input conditions:
\textit{Case 1:} $B(t)$ is a fundamental triangular signal ($E_1$);
    \textit{Case 2:} $B(t)$ is a third-harmonic triangular signal with a DC offset ($E_3 - E_0$);
    \textit{Case 3:} $B(t)$ is the superposition of the two previous cases: $E_1(t) + E_3(t) - E_0$.

To provide an explanation for this CCO phenomenon, they  normalized the aforesaid system, reducing its complexity to
\begin{equation}
\label{eq:dimensionless_sys}
\begin{cases}
\dot{x}=s, & \text{for } x<1 \\
x(\tau_{+})=b(\tau_{+}) & \text{if } x(\tau)=1
\end{cases}
\end{equation}
with $s = \frac{IT}{CV_{T}}$ being the normalized slope parameter and $b(\tau)$ is the normalized base signal so that for Case 3 
\begin{equation}
\label{eq:dimensionless_base}
b(\tau) = e_{1}(\tau)+e_{3}(\tau)-b_{0}
\end{equation}
where $b_{0} = E_{0}/V_{T}$ is the normalized DC offset.
This allows its dynamics to be governed by only two key parameters: $a$ and $b_0$, which act as a DC shift, positioning the orbits relative to the stability points. They found that the behavior over time is perfectly captured by a Peak Phase Return Map, $f$, which maps the timing of one spike, $\theta_n$, to the timing of the next, $\theta_{n+1}$, defined by $f(\theta_n) = F(\theta_n) \mod 1$. Due to the piecewise-linear nature of the input signals, this return map was exactly defined by nine distinct segments. Four out of these nine segments exhibit CCO behavior with the slope of that map, $Df_3(\theta)$,  exactly equal to zero.
The zero-slope segments -- where $d_1 = (a+1)/(4a)$ -- then define a constant output time
\begin{align*}
S_{1}: \tau_{n} &\in \left[\frac{1}{12},\frac{3}{12}\right], & F_{3}(\tau_{n}) &= \frac{2}{12}a+1+b_{0} \\
S_{2}: \tau_{n} &\in \left[d_{1},\frac{5}{12}\right], & F_{3}(\tau_{n}) &= \frac{2}{12}a+\frac{3}{2}+b_{0} \\
S_{3}: \tau_{n} &\in \left[\frac{7}{12},1-d_{1}\right], & F_{3}(\tau_{n}) &= -\frac{2}{12}a+\frac{3}{2}+b_{0} \\
S_{4}: \tau_{n} &\in \left[\frac{9}{12},\frac{11}{12}\right], & F_{3}(\tau_{n}) &= -\frac{2}{12}a+2+b_{0}.
\end{align*}

The central finding -- which unlocks the paradox -- is that while Case 1 and Case 2 independently drive the system into chaos, when the expansion parameter is sufficiently high -- namely $a > 2$ --, their combination in Case 3 leads to the profound stability of a superstable periodic orbit, SSPO.

The value $F_3(\tau_n)$ is constant within each $S_i$. This physical flattening means that regardless of the exact phase of the spike within $S_i$, the next spike occurs at the same fixed time. This represents an infinite compression of the phase space, making these segments the source of super-stability.

That being said, herein, the CCO phenomenon is the direct consequence of the battle between expansive chaos and super-stable compression. When $a > 2$, the return maps for the isolated inputs (Case 1 and 2) have slopes greater than one, leading to chaotic expansion; however, when combined (Case 3), $f_3$ displays both highly expansive regions -- where $|Df_3|=2a > 4$ -- and four infinitely compressive regions, $|Df_3|=0$. An orbit that begins in chaos is rapidly drawn to and captured by one of these zero-slope segments. Any trajectory that hits an $S_i$ is instantly funneled into an SSPO. 

The authors also perform a comprehensive mapping of the regions where SSPO exists in the $(a, b_0)$ parameter space. The condition for a period-$k$ SSPO is simply that the image of an $S_i$ interval returns to itself after $k$ steps: $f^k(f(S_i)) = f(S_i)$. This calculation revealed that the existence regions for SSPOs of periods 1 through 4 are remarkably complex, often non-contiguous, and highly sensitive to the parameters. The study also demonstrated that, for specific parameter settings, the system can exhibit coexisting SSPOs -- e.g., both a period-1 and a period-4 orbit --, highlighting the complexity of the underlying phase space.

\textbf{Deep Reinforcement Learning. --- }
As machine learning and artificial intelligence become ever more popular, it is worth asking to what extent Parrondian strategies can be employed therein. An application of chaos control within CCO was successfully implemented in Ref.~\cite{li2025chaos}, allowing leveraging the computational power of Deep Reinforcement Learning (DRL)~\cite{sutton1998reinforcement}. In the aforesaid work, Li, Li, and Miyoshi designed a specific neural-network-based function, $\mathcal{C}$, -- which they  trained using the Proximal Policy Optimization algorithm~\cite{raffin2021stable} -- to control the Lorenz system~\cite{lorenz2017deterministic},
\begin{align}
\dot{x} &= \sigma \, (y - x) \nonumber \\
\dot{y} &= x \, (\rho(t) - z) - y \label{eq:lorenz} \\
\dot{z} &= x \, y - \beta \, z. \nonumber
\end{align}
The controller $\mathcal{C}$ operates under a severe constraint: it can only apply a unidirectional perturbation to the chaos-controlling parameter ($\rho$); in other words, its action is limited to either doing nothing -- $\Delta\rho(t)=0$ -- or applying a small increase in $\rho$ of $1 \% $. Since increasing the parameter is known to maintain or even enhance the chaotic strength the Lorenz system in the region $\rho = 28$, the DRL agent must learn strategy to achieve order 
using a reward function,
\begin{equation}
\label{eq:reward}
R(t) = - || \mathbf{x}(t) - \mathbf{x}_{\text{target}} ||^2 + R_{\text{bonus}} \cdot \mathbb{I}(\mathbf{x}(t) \in \mathcal{R}),
\end{equation}
-- where $R_{\text{bonus}}$ is a positive constant reward for successfully entering the target region $\mathcal{R}$, and $||\cdot||$ is the Euclidean norm -- that encourages stabilization towards a small region $\mathcal{R}$ near the target state $(\bar{x}, \bar{y}, \bar{z})$ corresponding to a desired periodic orbit by maximizing the cumulative reward. 

Overall, the controlled system alternates between two chaotic states -- default chaos and the temporarily enhanced chaos --, yet the overall combined dynamics leads to the suppression of chaos and the establishment of stable order, with the DRL agent providing a practical, high-dimensional example of the `Chaos + Chaos = Order' paradox.
Unlike traditional control -- eg, Ott-Grebogi-Yorke~\cite{ott1990controlling} and adaptive~\cite{boccaletti2000control} -- methods to stabilize an existing unstable periodic orbit, this paradoxical switching strategy achieves control by creating a new stable periodic orbit in the Lorenz system where none existed before. Moreover, this work powerfully validates the chaotic  PP  as a viable control mechanism and establishes Machine Learning as an effective tool for discovering such nonlinear  control strategies.

\subsection{Engineering}

\textbf{Secure Communication. --- } The CCO principle has been applied to secure communication systems. 
For instance, using the Chen chaotic system~\cite{kumar2018chaotic} explored the secure transmission of RGB images simulated in Matlab-Simulink. 
The Chen system~\cite{chen1999controlling} is a three-dimensional chaotic dynamical system described by the following equations,
\begin{align}
\label{eq:chen}
\begin{cases}
\frac{dx}{dt}=a(y-x) \\
\frac{dy}{dt}=(c-a)x-xz+cy \\
\frac{dz}{dt}=xy-bz;
\end{cases}
\end{align}
the parameters of the system --- $a$, $b$, and $c$ --- are positive real constants, with values of 35, 3, and 28, respectively. Chaos in this system can be controlled by changing $c$ which assumes the role of control parameter in their protocol. According to its bifurcation diagram, the Chen system shows that it is chaotic over a wide range of $c$, except at a specific value of $c^* = 26.083$, where chaos is suppressed.
Nevertheless, Kumar et al. introduce a theorem proving that a stabilized Chen attractor can also be synthesized by using the average value of the parameters, $p^*$,
    \begin{align}
    p^{*}\equiv\frac{\Sigma_{l=1}^{N}  w_{i}\, c_{i}}{\Sigma_{i=1}^{N}w_{i}},
    \end{align}
where $w_i$ are the weights associated with each parameter. This corresponds to an interpretation of the  PP  in the CCO form that the authors employ, considering a master-slave topology afterwards. Explicitly, chaos is added to the image at the transmitter end -- the master --  for encryption and then removed at the receiver end -- the slave -- by stabilizing the system, thereby retrieving the original image with minimal loss of information. In this way, they propose the stabilization of a chaotic system by alternating between two sets of control parameters, each of which would individually produce chaotic behavior. 
These authors compare the correlation factor between the original and received images to assess the effectiveness of the proposed approach.
The method presents as being quite effective as it has a full correlation between the original and received images, whereas it is able to secure a correlation between the original and chaotic transmitted images as low as $0.035$.

This concept was physically implemented by~\cite{adeyemi2021fpga} using a Field-Programmable Gate Array (FPGA) -- i.e., an integrated circuit that can be configured by the user to perform a specific function after it has been manufactured --  translating and confirming the results of the theoretical concept of using parameter switching to control chaos for secure image transmission into a practical, real-world hardware application.

\textbf{Controllability of Optical Chaos. --- }
The inability to observe the temporal dynamics of optical chaos due to the ultrafast propagation of light~\cite{gao2014single} is a long-standing challenge in nonlinear  optics. Previous investigations relied exclusively on static imaging or spectral measurements, which capture only time-averaged phenomena and fail to reveal the dynamic evolution and critical sensitivity that define chaos~\cite{volos2012chaotic,wang1991ballistic}. This means that the capability to reveal and control real-time chaotic dynamics is critical for further understanding and engineering these systems for applications like optical communication and cryptography~\cite{uchida2008fast,ploschner2015seeing}. An alternative approach was introduced by~\cite{fan2021real} in which they deal with the single-shot, real-time spatial-temporal imaging of an optical chaotic system using compressed ultrafast photography (CUP)~\cite{garcia2001spatiotemporal,altmann2013leaking,davis1981quantum}. This advance overcomes the intrinsic nonrepeatability constraint of chaotic events, which makes traditional ultrafast imaging techniques requiring repeated measurements unsuitable.
The CUP system captures light from the optical cavity placed at its object plane~\cite{chinnery1996experimental}, and then a femtosecond laser introduces the pulse jointly with some water -- which creates weak optical scattering -- to allow the camera to register the motion of light. 
After the image acquisition and reconstruction phase, the image is encoded by a digital micromirror device with a pseudorandom binary pattern, which is then temporally sheared by a sweeping electric field inside a streak camera before being integrated and captured by a charge-coupled device (CCD) detector in a single exposure~\cite{kellert1993wake,haydn2005hitting}.
This process yields single-shot snapshots, revealing the time evolution of the system's phase map without the need for repetition~\cite{monifi2016optomechanically}, which permitted the experimental verification of extreme sensitivity to initial conditions -- the core feature of chaos -- in real time.
The ability to simultaneously control and monitor optical chaotic systems was demonstrated using a Kerr gate within a quarter Bunimovich stadium cavity~\cite{fang2005analysis,backer2008dynamical}. This result provides an analogy for Parrondian approaches in dynamical systems, where two `non-losing' processes in sequence can yield a surprisingly different, potentially more powerful, result.
Using a Kerr gate within a quarter Bunimovich stadium cavity the system is capable of simultaneously controlling and monitoring optical chaos. This instance provides an analogy for Parrondian approaches in dynamical systems, where two `non-losing' processes in sequence -- closing (regular trajectory) $\rightarrow $ opening  (chaotic trajectory) the Kerr gate -- can yield a surprisingly different, potentially more powerful, result. The observed $\text{Regular Mode} \rightarrow \text{Chaotic} \rightarrow \text{Regular } \ldots $ sequence supports the concept that temporary or controlled exposure to a chaotic mechanism can shift the final stable state of a system.
In the context of  PP, where the sequence of two processes can yield an outcome greater than their simple sum, this demonstrates a form of \textit{Order + (Controlled) Chaos = New Order}. The ability to use chaos not as an uncontrollable instability, but as a transitional catalyst, paves the way for a deeper engineering of chaotic systems~\cite{alt1999experimental,sciamanna2015physics}.
By offering temporal and spatial information in a single shot, the CUP technique opens new avenues for studying complex phenomena, such as Poincaré recurrence time and dynamic tunnelling between modes, which have been previously limited by time-integrating methods~\cite{haydn2005hitting}.

\textbf{Optimization Algorithms. --- } 
Optimization problems are more often characterized by complexity, multimodality and discontinuity, which require dedicated algorithms like those classified as metaheuristic, such as Genetic~\cite{gandomi2013metaheuristic}, and Particle Swarm Optimization~\cite{derrac2011practical} as well as Gravitational Search~\cite{mirjalili2017gravitational}, GSA, and Biogeography-Based Optimization~\cite{simon2008biogeography}, BBO.
Chaotic maps -- the inclusion of which significantly increases the performance of these algorithms in comparison to random perturbations -- have been used to replace fixed parameters in these algorithms to increase population diversity and mixing capability, which helps avoid premature convergence. 

Nonetheless, in Refs.~\cite{kumar2022alternated,rani2019alternated}, the authors showed the alternation between two chaotic dynamics -- very much in the spirit of  PP  -- can further enhance optimization mechanisms in metaheuristic algorithms by improving the precision and ability to find global solutions; specifically, the authors assert that when using sequences generated by alternating two logistic maps the system can be conducted to a superior orbit, superior in the sense that is ideal for achieving higher optimization rates.

\subsection{Economics and Social Systems}
The paradox holds valuable implications for financial models and investment strategies. It highlights the importance of diversification and risk management, challenging the conventional wisdom that success comes from consistently selecting winning investments.

\textbf{Duopoly models. --- } Duopolies are markets where two firms compete to produce similar or perfect
substitute goods. In some cases, the optimization of the utility functions
gives rise to solving the system%
\begin{equation*}
\left\{
\begin{array}{c}
x=f(y), \\
y=g(x),%
\end{array}%
\right.
\end{equation*}%
to obtain the equilibrium points (so-called Cournot points), where $x$ and $y
$ are the outputs of each firm. The functions $f$ and $g$ are called
reaction functions. Then, under naive expectations, the production of both
maps are planned as
\begin{equation*}
\left\{
\begin{array}{c}
x_{n+1}=f(y_{n}), \\
y_{n+1}=g(x_{n}),%
\end{array}%
\right.
\end{equation*}%
where $x_{n}$ and $y_{n}$ are the outputs of both firms at time $n$. So, the market can be modeled by a map of the form $T(x,y)=(f(y),g(x))$. We refer the reader to~\cite{puu},~\cite{norin}, or~\cite{kopel} for examples of these models. Note that the second iterate of the map $T$ has the form%
\begin{equation*}
T^{2}(x,y)=\left( (f\circ g)(x),(g\circ f)(y)\right) ,
\end{equation*}%
and hence the dynamics depend on the composition of both maps. Thus, the
Parrondo paradox can be seen in terms of the reaction functions, which can
be simple while the dynamics of the duopoly is chaotic. For instance, in the famous Puu's duopoly~\cite{puu}, the reaction functions are given by%
\begin{equation*}
f(y)=\sqrt{\frac{y}{c_{1}}}-y\text{ and }g(x)=\sqrt{\frac{x}{c_{2}}}-x,
\end{equation*}%
where $0<c_{1}\leq c_{2}$ are the constant marginal cost of each company. It is easy to see that the dynamics of $f$ and $g$ is simple, while Puu's
doupoly becomes chaotic when $c_{1}/c_{2}>6.18...$~~\cite{Canovas3}, and so we have an example of "simple+simple$\to$chaos" paradox that enriches the dynamics of the model.

\textbf{Control of traffic systems. --- }
The paradoxical combination of system behaviors can also be verified in macroscopic examples where the challenge is to control inherently unpredictable, real-world dynamics. A compelling illustration of this principle is found in the field of control engineering, as implemented in Ref.~\cite{kumari2020novel} for traffic flow control modelled
with the logistic function $f_{a}(x)=a \,x \,(1-x)$ was introduced. In this context, the variable $x$ represents the normalizsed traffic density on a stretch of road, while the parameter $a$ relates to factors like road capacity or the rate at which congestion propagates. When $a$ is high, the model exhibits chaotic oscillations in traffic density, simulating severe congestion and unpredictable stop-and-go patterns. The application of the chaotic Parrondo scheme is therefore a direct attempt to steer the traffic density from these unstable chaotic states into a predictable, ordered flow.
The work by Kumari and Chugh leverages the principles of the chaotic Parrondo paradox by constructing a controlled system from a sequence of three individually non-chaotic steps. Each step is based on a modified form of the chaotic logistic function achieved via a SP-iteration~\cite{phuengrattana2011rate}
\begin{equation}
f_{i}(x)=(1-\lambda _{i})x+\lambda _{i}f_{a}(x), \quad \lambda _{i}\in (0,1) \text{ for } i=1,2,3.
\end{equation}
In this formulation, $\lambda _{i}$ functions as the control parameter or weighting factor of the model. When $\lambda_{i}$ is small, the map $f_{i}$ places a substantial weight on the current state $x$, effectively damping the high-frequency, chaotic expansion dictated by the original map $f_a(x)$. For specific values of $\lambda_i$, the individual map $f_{i}$ successfully achieves stabilization, yielding an ordered, non-chaotic outcome.
The overall controlled system is then defined by the sequential composition of these three steps: $F(x) = f_{1}\circ f_{2}\circ f_{3}$.

A striking result -- elaborated upon in~\cite{canovas2021dynamics} -- is the demonstration of the OOOC chaotic Parrondo paradox.
More precisely, the control parameters $\lambda _{i}$ can be chosen such that each individual controlled map $f_{i}$ is regular, showing a vanishing topological entropy, $s(f_{i})=0$. Topological entropy is a measure of the exponential complexity of the orbits, and zero entropy signifies a highly predictable, non-chaotic state. In spite of that, the sequential iteration of these three simple processes, $F=f_{1}\circ f_{2}\circ f_{3}$, generates a composite map with positive topological entropy ($h(F)>0$). The control mechanism -- i.e., the combination of the simple operations -- has therefore introduced the very chaotic instability it was designed to suppress. 

For real-world applications, this topological paradox is demonstrated to be physically observable as a chaotic paradox. 
The authors demonstrate that the positive topological entropy, $h(F)>0$, implies that the maximum Lyapunov exponent for some of the turning points of the composite map $F$ is also positive. This indicates that the system trajectories become exponentially sensitive to initial conditions under combined control, confirming the practical manifestation of chaos.

Other examples of the emergence of chaotic Parrondian effects can be found in biological and ecological systems. Hereinafter, we list some custom cases.

\subsection{Biologic and Ecological}

\textbf{Chaotic ecological delayed systems. --- }
The paradigmatic example of manipulating chaoticity from a parrondian perspective in delayed systems is the delayed logistic equation~\cite{mendoza2018parrondian},
\begin{align}
X_{n+1}&=Y_{n} \label{eq:1} \\
Y_{n+1}&=C Y_{n}(1-X_{n}) , \label{eq:2}
\end{align}
for which chaotic dynamics are typically seen for values of $C$ greater than $2.0$. The authors introduce a switching strategy in which the value of the parameter $C$ alternates between even and odd values, $C_e$ and $C_o$, at each iteration. With that, they can obtain examples of the CCO phenomenon. For instance, using $C_o=2.10$, a region of periodic oscillations is found for $C_e$ values from 2.26 to 2.27, even though both parameters on their own would result in chaotic behavior. Bearing in mind the ecological origin of the logistic equation, the authors apply the same reasoning in other two-dimensional maps used in ecology and demographics, namely the Lotka-Volterra~\cite{lotka1925elements,volterra1926variazioni}, Ricker~\cite{ricker1954stock}, and Beddington~\cite{beddington1975dynamic} maps. They show that the OOC  inverse effect, where two stable periodic behaviors are combined to create a chaotic outcome, is also achievable for these cases. These results represent an important case of the stabilization of chaotic dynamics for models of seasonality in ecology.

\textbf{Seasonality in ecological models. ---} In ecological models, seasonality refers to the cyclical and predictable variations of environmental factors. On the other hand, the combination of two unfavorable conditions, such as winter and summer, can generate stable and desirable behavior, even if individually these factors lead to chaos or extinction, giving rise to Parrondo's paradox. In Ref.~\cite{yadav2019seasonality} , the authors apply the alternation strategy (Eq.~\ref{eq:48}) using a superior logistic model.

\begin{equation}
    \left ( B_{1},B_{2} \right ):\left\{\begin{matrix}
x_{n+1}=x_{n}^{2}+c_{1} & when & n & is & odd \\
x_{n+1}=x_{n}^{2}+c_{2} & when & n & is & even \\
\end{matrix}\right.
\label{eq:48}
\end{equation}

\noindent where $B_{1}$ and $B_{2}$ are two different discrete dynamics, and $x$, $c$, $c_{1}$, $c_{2} \in \mathbb{R}$.

The authors observed through the bifurcation diagram that a population heading towards extinction can enter a stable oscillation after the transition, that is, "Extinction + Undesirable $\to$ Desirable".

\textbf{Fish school control. ---}
Another example of ecological modelling in which assuming a simple form of seasonality by alternating (two) different parameter values is found in the Modified Beverton-Holt Model introduced by~\cite{mendoza2018switching} as well. The Beverton-Holt map~\cite{beverton2012dynamics}, 
\begin{align}
X_{n+1}=\frac{C~X_{n}\, \exp[-X_{n}]}{1+b(1-\exp[-X_{n}])},
\label{eq:1}
\end{align}
which is frequently used to model fish populations, even in the fisheries business. The model is mostly ruled by the parameter $C$ with larger values of $b$ tend to stabilize the dynamics. Therefore, considering different values of $b$ to identify the parameter values associated with undesirable behavior, such as extinction or chaos, the authors introduced a switching strategy where the parameter $C$ is alternated between an even iteration parameter, $C_e$, and an odd iteration parameter, $C_o$. With this switching strategy it is possible to achieve stable oscillations even when the individual parameter values would lead to chaotic or extinction dynamics. The bifurcation diagrams in the paper illustrate how the switching strategy creates regions of stable, periodic behavior where chaos would normally be expected. For example, it was noticed that when $b=1/2$ and $C_o = 29$, which is a chaotic pair of parameters, stable oscillations can be observed for $C_e$ values between 36.25 and 37.

\textbf{Generalized Beverton-Holt map. --- }
In~\cite{de2013prediction} the authors study the stabilization of an unstable periodic orbit (UPO) through a periodic prediction-based control
(PBC). 
While stabilizing the UPO, the use of control intervention at every step (non-periodic PBC) might introduce the undesirable Allee effect into the system dynamics. The study key finding is that switching to a periodic control (intervention only at certain time intervals) can solve this drawback, weakening the induced Allee effect and the associated high extinction risk.


\textbf{Host-parasitoid dynamics. --- }
Another case emerging from quantitative Biology concerns the Beddington, Free, and Lawton (BFL) model used to study host-parasitoid interactions~\cite{nicholson1935balance}. It is governed by
\begin{align}
X_{n+1}=X_{n}\, \exp[r(1-X_{n})-Y_{n}]=f_{r}(X_{n},Y_{n})
\label{eq:1}
\end{align}
\begin{align}
Y_{n+1}=cX_{n}(1-\exp[-Y_{n}])=g_{c}(X_{n},Y_{n}).
\label{eq:2}
\end{align}

Similarly to other foregoing cases, the authors of Ref.~\cite{mendoza2018parrondo} focused on the parasitoid growth parameter, $c$ while fixing the host growth parameter at $r=2.0$.  For $Y=0$, the BFL case reduces to the Ricker model, which is used to identify parameter values that individually lead to chaotic orbits. By applying a switching strategy that alternates the parameter $c$ between two values, $c_o$ and $c_e$, for odd and even iterations, respectively, it was shown that chaotic dynamics can be transformed into stable periodic orbits. In addition, they assert the discover of two new parrondian desirable dynamic situations: ``quasiperiodic + quasiperiodic = periodic'' and `chaos + chaos = periodic coexistence,'' where the alternation of chaotic dynamics yields two different coexisting periodic stable orbits.

\section{Open questions}

In the last section, we draw some open problems regarding the topics discussed in this review.

\subsection{When the paradox is not possible}

Here, we wonder about when the paradox is not possible. So, we consider the triangular map $T:\Sigma _{2}\times X\rightarrow \Sigma _{2}\times X$ and two maps $f_{1},f_{2}:X\rightarrow X$ that are chaotic (resp. simple) and wonder on whether the dynamics of the projection $\pi _{2}\circ T$ is chaotic (resp. simple) in the phase space $X$. As we stated before~\cite{canovas2010analyzing}, if $f_{1}$ and $f_{2}$ are commuting piecewise monotone continuous interval maps with zero topological entropy, then the topological entropy of any sequence of maps containing only these two maps is zero as well. 

We wonder about these types of results under more general assumptions as well as for specific parameter-dependent models. Taking into account Table \ref{tab:parrondo_types} for the logistic family, we know from~\cite{canovas2013revisiting} that the physically observable paradox is possible, but the topological paradox has been proved for the case OOC. It is unknown whether the topological paradox CCO is possible for this family, or for other models. The implications for the physical Parrondo paradox CCO are clear; we observe periodic orbits after combining two chaotic maps because the chaotic set is hidden inside a set of zero Lebesgue measure. 
It must be recalled that the topological paradox CCO is possible, but under the hypothesis of~\cite{canovas2006dynamic}.

\subsection{The number of maps}

As mentioned throughout this survey, it seems that the paradox
"simple+simple$\to$chaos" is more likely if we iterate more than two maps, or
iterate two maps periodically with a higher period. To illustrate this problem, let us consider the logistic family $f_{a}(x)=ax(1-x)$ and choose two parameters $a_{1},a_{2}$ such that both maps are simple. The question is whether a periodic iteration of these maps can be chaotic or not. How is the parameter set $(a_{1}, a_{2})$ such that the Parrondo paradox takes place and when it is not possible?

\subsection{High-dimensional maps and flows}

Although there are some exceptions, most papers on the dynamic Parrondo paradox are set in the context of one-dimensional maps. There is a natural reason for this fact; namely, the dynamics of smooth enough one-dimensional maps is quite well known, but the same does not happen for high-dimensional maps. We expect the paradox to exist in higher-dimensional systems, but so far it is unclear how it manifests. To start with, it will be helpful to analyze the presence of the paradox in the order-two logistic families given by the difference equations $x_{n+1}=ax_{n}(1-x_{n-1})$ and $x_{n+1}=bx_{n-1}(1-x_{n})$, for suitable parameters $a$ and $b$. Another families with deserve investigation are H\'{e}non map~\cite{henon}
\begin{equation*}
\left\{
\begin{array}{l}
x_{n+1}=1-ax_{n}^{2}+y_{n}, \\
y_{n+1}=bx_{n},%
\end{array}%
\right.
\end{equation*}
and Lozi map~\cite{lozi}
\begin{equation*}
\left\{
\begin{array}{l}
x_{n+1}=1-a|x_{n}|+y_{n}, \\
y_{n+1}=bx_{n}.%
\end{array}%
\right.
\end{equation*}

A similar question can be posed for continuous-time models. Although Filippov systems had been studied by several authors~\cite{biak}, we emphasize that the change of the field is produced through a frontier in the phase space, and not on time. The approach of~\cite{cat3}, where the change is produced after a period of time is the proper paradox which may deserve further explorations.

\subsection{Parrondo’s paradox and the Allee effect}
As discussed with the VOSviewer and in the section on interdisciplinary analysis, there are several instances of the PP in ecology. Despite that, there is no systematic investigation of how the intensity of the Allee Effect impacts the frequency of occurrence of the PP. This effect 
plays an important role in 
population ecology~\cite{taylor2005allee,berec2008models,sun2016mathematical,pires2022randomness} and metapopulations~\cite{amarasekare1998allee,pires2019optimal} both from a theoretical or empirical side~\cite{kramer2009evidence,fauvergue2013review,angulo2018allee,branco2024widespread}. 
In this line of research, it will be important to develop exactly solvable biological models~\cite{petrovskii2005exactly}.

\subsection{Hybrid paradox}

Here, we wonder about the paradox when the dynamics involved are not homogeneous. For instance, let us consider a second-order difference equation of the form $x_{n+1}=f(x_{n},x_{n-1})$ and a one-dimensional system $x_{n+1}=g(x_{n})$ and wonder about the existence of the paradox when we iterate them alternately. We might also consider a continuous flow which is applied in times $[2n,2n+1]$, $n\in \mathbb{N}$, and a discrete dynamical system which we iterate from $2n+1$ to $2n+2$. Can we generate complex dynamics from simple generators and vice versa?

\subsection{Experimental works }

Several chaotic maps have been experimentally verified, such as 1D models: logistic map~\cite{suneel2006electronic}, tent map~\cite{campos2009simple} and combined logistic-tent maps~\cite{garcia2013difference}. Experimental implementations have also been possible for coupled-logistic maps~\cite{l2016electronically,mhiri2019experimental} and a network of maps~\cite{gutierrez2025non}.

For classical systems, the authors of~\cite{kirikawa2012bifurcating} devised an electronic circuit implementation of the paradox where they use a plural bifurcating neuron (BN) system to construct pulse-coupled neurons (PCN), and they experimentally confirm both facets of the PP: Chaos + Chaos $\to$ Order (CCO), and Order + Order $\to$ Chaos (OOC). %

In another study involving PCNs~\cite{inagaki2007response}, the behavior of a chaotic spiking circuit (CSC) was evaluated when it receives periodic and non-periodic impulses. The experiment studies how the alternating application of these inputs to a chaotic spiking neuron can induce order through complexity. 

For quantum implementations, the chaotic version of the PP awaits experimental realization, though experimental advances 
have been established for non-chaotic quantum walks in photonic platforms~\cite{jan2020experimental}.
 
\subsection{Parrondo’s paradox and different routes to chaos}

Despite the extensive analysis of PP within chaotic systems, a systematic investigation of its interplay with the fundamental transition mechanisms to chaos is notably absent. Specifically, is the manifestation of the paradox dependent on the route by which chaos is achieved? This question is crucial, as classical pathways like period-doubling, intermittency, and quasiperiodicity impact distinct signatures on a system's dynamics and bifurcation structure. This question could be addressed in diverse ways. One approach is to employ generalized models, such as~\cite{pires2025composing}, capable of exhibiting diverse routes to chaos (gradual and abrupt) through parameter variation.

\section{Final remarks}

This review provides  the first systematic synthesis of the intricate relationship between Parrondo's paradox and chaos. 

We conducted a comparative analysis that distinguishes observable from topological chaos, elucidated the bidirectional PP-chaos connection for both protocol design and instability control, and demonstrated the paradox's ubiquitous interdisciplinary nature. 
We provide a Python code~\cite{codes_pp_chaos}  that enables researchers to directly engage with these phenomena.

For future advances, 
we note that increased efforts will be important to promote more bridges between experimental and theoretical aspects of the PP.

\section*{Acknowledgments}

S.M.D.Q. thanks CNPq (Grant No. 302348/2022-0) and FAPERJ (SEI-260003/005741/2024) for financial support.
M.A.P. acknowledges financial support from CNPq (Grant No. 310093/2025-2).

\bibliography{main.bib}

\appendix

\section{Additional technical details about dynamical systems}\label{app:chaos_details}

In this appendix we present a practical characterization of chaos for regular enough interval maps. A set $\Omega \subset X$ is called a forward invariant under $f$ if $f(\Omega )=\Omega $. The basin of attraction of a forward invariant set $\Omega $, $B(\Omega )$ is given by
\begin{equation*}
B(\Omega )=\{x:\omega (x,f)\subset \Omega \}.
\end{equation*}%
A forward invariant set $\Omega$ is defined as a (minimal) metric attractor if its basin of attraction $B(\Omega)$ has positive Lebesgue measure, and if any proper, forward invariant, compact subset $\Omega ^\prime \subset \Omega$ has a basin $B(\Omega ^\prime)$ with Lebesgue measure equal to zero. This definition, originating with Milnor~\cite{Milnor}, establishes $B(\Omega)$ as the basin of the attractor. For some smooth enough interval maps, the number of these Milnor attractors can be precisely characterized~\cite{vanstrien}. We will assume that the map $f$ is smooth enough, in our case $C^{3}$, and the turning points are non-flat, that is, for $x$ close to $c_{j}$, $j=1,2,...,i-1$,
\begin{equation*}
f(x)=\pm |\phi (x)|^{\beta _{i}}+f(c_{j}),
\end{equation*}%
where $\phi $ is $C^{3}$, $\phi (c_{j})=0$ and $\beta _{i}>0$. An interval $J\subset X$ is called wandering if the intervals $\{J,f(J),\ldots ,f^{n}(J),\ldots \}$ are pairwise disjoint and if $J$ is not contained in the basin of an attracting periodic orbit of $f$. The following result can be found in~\cite{vanstrien}.

\begin{theorem}
\label{theowan}Let $f:X\rightarrow X$ be a $C^{2}$ piecewise monotone map
with non flat turning points. Then $f$ has no wandering intervals.
\end{theorem}

In addition, if the map $f$ is $C^{3}$ and piecewise monotone there are
three possibilities for its metric attractors:

\begin{itemize}
\item[A1)] A periodic orbit (recall that $x$ is periodic if $f^{n}(x)=x$ for
some $n\in \mathbb{N}).$

\item[A2)] A union of periodic intervals $J_{1},...,J_{k}$, such that $%
f^{k}(J_{i})=J_{i}$ and $f^{k}(J_{i})=J_{j}$, $1\leq i<j\leq k$, and such
that $f^{k}$ is topologically mixing. The topologically mixing property implies the existence of dense orbits on each periodic interval (under the iteration of $f^{k}$).

\item[A3)] A minimal (every orbit is dense) Cantor set $\omega (c,f)$ such
that $c\in \omega (c,f)$ is a turning point.
\end{itemize}

Attractors of types (A2) and (A3) must include the orbit of a turning point. Consequently, the total number of such attractors is limited by the number of turning points of the function $f$.
Furthermore, consider any critical point $c$ that does not belong to the basin of a periodic attractor. For such a point, there exists a neighborhood $U$ with the following property: if an iterate $f^{n}(x)$ lands within $U$ for any point $x$ and iteration $n \in \mathbb{N}$, then the Schwarzian derivative of the next iterate, $f^{n+1}$, evaluated at $x$, must be negative, i.e., $\mathcal{S}(f^{n+1})(x)<0$. Recall that the Schwarzian derivative of a $C^{3} $ map is given by%
\begin{equation*}
\mathcal{S}(f)(x)=\frac{f^{3)}(x)}{f^{\prime }(x)}-\frac{3}{2}\left( \frac{%
f^{\prime \prime }(x)}{f^{\prime }(x)}\right) ^{2},
\end{equation*}%
for all $x\in \lbrack 0,1]$ such that $f^{\prime }(x)\neq 0$. If we add the condition of a negative Schwarzian derivative, then we have the following result that can be found in~\cite{Singer}.

\begin{theorem}
\label{attractors} Let $f:I\rightarrow I$ be a $C^{3}$ interval map with $\mathcal{S}(f)(x)<0$ for each $x\in \lbrack 0,1]$ such that $f^{\prime }(x)\neq 0$. Then each attracting periodic orbit attracts at least one critical point or boundary point.
\end{theorem}

\FloatBarrier

\end{document}